\documentclass[sigconf, nonacm]{acmart}
\usepackage{algorithm}
\usepackage{algpseudocode}
\usepackage{xspace}
\usepackage{threeparttable}
\usepackage{color}
\usepackage{url}
\usepackage{caption}
\usepackage{subcaption}
\usepackage{paralist}
\usepackage{wrapfig}
\usepackage{multirow}
\usepackage{listings}
\usepackage{comment}
\usepackage{booktabs}
\usepackage{makecell}
\usepackage{boxedminipage}
\usepackage{amsmath}
\usepackage{graphicx}
\usepackage{minibox}
\usepackage{pifont}
\usepackage{color}
\usepackage{subfiles}
\usepackage[utf8]{inputenc}
\usepackage[english]{babel}



\newcommand{\bheading}[1]{{\vspace{4pt}\noindent{\textbf{#1}}}}
\newcommand{\iheading}[1]{{\vspace{1pt}\noindent{\textit{#1}}}}
\newcolumntype{?}{!{\vrule width 1pt}}

\newcounter{note}[section]

\newcommand{\ssecref}[1]{\mbox{\S\ref{#1}}\xspace}

\newcommand{\ignore}[1]{}

\newcommand{\ie}{\textit{i.e.}\xspace}
\newcommand{\eg}{\textit{e.g.}\xspace}

\newcommand{\etal}{\textit{et al.}\xspace}
\newcommand{\resp}{\textit{resp.}\xspace}

\newcommand{\framework}{\textsc{EBFT}\xspace}
\newcommand{\sysnameSyn}{\textsc{EBFT-Syn}\xspace}
\newcommand{\sysnamePSyn}{\textsc{EBFT-PSyn}\xspace}
\newcommand{\sysname}{\textsc{EBFT-Turbo}\xspace}

\newcommand{\comVote}{\textsf{comVote}\xspace}
\newcommand{\witVote}{\textsf{witVote}\xspace}
\newcommand{\comVotes}{\textsf{comVote}s\xspace}
\newcommand{\witVotes}{\textsf{witVote}s\xspace}

\newcommand{\kbytes}{\ensuremath{\mathrm{KB}}\xspace}

\newcommand{\secs}{\ensuremath{\mathrm{s}}\xspace}

\newcommand{\cmark}{\ding{51}}%
\newcommand{\xmark}{\ding{55}}%

\newcounter{packednmbr}

\newenvironment{packeditemize}{
    \begin{list}{$\bullet$}{
        \setlength{\labelwidth}{0pt}
        \setlength{\itemsep}{2pt}
              \setlength{\leftmargin}{\labelwidth}
              \addtolength{\leftmargin}{\labelsep}
              \setlength{\parindent}{0pt}
              \setlength{\listparindent}{\parindent}
              \setlength{\parsep}{1pt}
              \setlength{\topsep}{1pt}}}{\end{list}}

\newtheorem{theorem}{Theorem}
\newtheorem{lemma}{Lemma}

\newtheorem{remark}{Remark}
\newtheorem{definition}{Definition}
\newcommand{\e}[1]{{\mathbb E}\left[ #1 \right]}

\setcopyright{none}
\renewcommand\footnotetextcopyrightpermission[1]{} 

\newcommand\vldbdoi{XX.XX/XXX.XX}
\newcommand\vldbpages{XXX-XXX}
\newcommand\vldbvolume{14}
\newcommand\vldbissue{1}
\newcommand\vldbyear{2020}
\newcommand\vldbauthors{\authors}
\newcommand\vldbtitle{\shorttitle} 
\newcommand\vldbavailabilityurl{URL_TO_YOUR_ARTIFACTS}
\newcommand\vldbpagestyle{plain} 
\definecolor{red}{RGB}{215,135,97}

\begin{document}

\title{\framework: Simplifying BFT Consensus Through Egalitarianism}

\author{{Jianyu Niu$^{1}$$^\dagger$, Runchao Han$^2$$^\dagger$, Shengqi Liu$^{1}$, Fangyu Gai$^{3}$, Ivan Beschastnikh$^4$, Yinqian~Zhang$^1$,~Chen~Feng$^3$}}
\affiliation{\institution{$^1$Southern University of Science and Technology \hspace{0.3cm} $^2$Monash University \& Data61 \\ University of British Columbia ($^3$Okanagan Campus, $^4$Vancouver Campus)}}
\affiliation{
    $^1$\{niujy@sustech.edu.cn,  liusq2020@mail.sustech.edu.cn, yinqianz@acm.org\} \hspace{0.3cm} $^2$runchao.han@monash.edu \hspace{0.3cm} $^3$\{fangyu.gai, chen.feng\}@ubc.ca  $^4$bestchai@cs.ubc.ca
}
\thanks{$\dagger$These authors have contributed equally to this work.}







\begin{abstract}
    We present Egalitarian BFT (\framework), a simple and high-performance framework of BFT consensus protocols for decentralized systems like blockchains.
    The key innovation in \framework is \textit{egalitarian} block generation: nodes randomly and non-interactively propose blocks containing client transactions, rather than relying on a leader to do so.
    Apart from deterministic safety and liveness guarantees standard in BFT protocols, the egalitarian design provides two novel features: (i) \framework is resilient to attacks targeting the leader, such as bribery and targeted DoS attacks, and (ii) \framework does not require any fail-over protocol to detect and replace the faulty leader.
    \framework consists of three protocols: \sysnameSyn for synchronous networks, \sysnamePSyn for partially synchronous networks, and \sysname that builds on \framework for high performance.

    We implement \framework and evaluate its performance on AWS.
    To compare \framework with state-of-the-art BFT protocols, we build \sysnamePSyn based on Bamboo, an open-source platform for prototyping partially synchronous BFT protocols.
    We evaluate \sysnamePSyn and HotStuff on EC2  with up to 16 nodes.
    The evaluation shows that \sysnamePSyn achieves better throughput and latency than HotStuff.
    To demonstrate its simplicity and practicality, we build \framework on the Go version of Bitcoin, \texttt{btcd}.
    We implemented \sysnameSyn, \sysnamePSyn and \sysname in about 920 LoCs in total. This indicates that \framework can be built on top of existing blockchains with relatively little effort.
    We evaluate these protocols on EC2 instances with up to 256 nodes.
    Our evaluation shows that \sysnameSyn (\resp \sysnamePSyn) achieves a latency of $6$ (\resp $1$) seconds, and an optimized version of \sysnamePSyn processes up to $3.6$k transactions per second and has a latency of $8$ seconds.
\end{abstract}

\maketitle

\pagestyle{\vldbpagestyle}

\section{Introduction} \label{sec:introduction}
Byzantine Fault Tolerant (BFT) consensus (also known as Byzantine State Machine Replication, SMR) enables a set of nodes to maintain a consistent ledger with a sequence of transactions, even in the presence of Byzantine nodes that behave arbitrarily.
As an important primitive in distributed computing, BFT consensus has been extensively studied~\cite{lamport82byzantinegeneral, HQReplica, 700BFT}, and many elegant protocols like PBFT~\cite{castro1999practical} and Zyzzyva~\cite{kotla2007zyzzyva} were proposed.
BFT has gained renewed interest due to its important role in building decentralized systems, especially blockchains~\cite{Algorand, BFTlens, suvBFT}.

\subsection{Existing Consensus Protocols}
However, the first-generation blockchains, such as Bitcoin~\cite{nakamoto2012bitcoin} and Ethereum~\cite{wood2014ethereum}, usually use Nakamoto-style consensus, rather than the well-studied classic BFT consensus protocols.
We believe there are two reasons for why developers of decentralized systems, like public blockchains, prefer Nakamoto-style consensus over classic BFT consensus:
1) the leaderless design in Nakamoto-style consensus leads to resilience against attacks on leaders; and
2) the overall design of Nakamoto-style consensus is much simpler than classic BFT protocols.

\vspace{1em}
\noindent \textbf{BFT protocols lack resilience against attacks on leaders.} In decentralized applications, if an adversary can predict the next leader node in a consensus protocol, then they can attack this node to slow down the protocol, or even break the protocol's security properties.
Such attacks include \emph{bribery attacks}~\cite{bonneau2016buy} where the adversary bribes the leader to vote on a certain block, and \emph{targeted denial of services (DoS) attacks}~\cite{chen2022efficient} where the adversary floods the leader with messages to make it unavailable to participate in the consensus protocol.

Classic BFT protocols use a \textit{stable leader} approach: a known and fixed leader prepares proposals and coordinates with other nodes to reach consensus~\cite{castro1999practical, kotla2007zyzzyva, HQReplica, 700BFT}.
This allows the adversary to launch attacks on this stable leader.
For example, a targeted DoS attack~\cite{chen2022efficient} on the leader can break the protocol's liveness.

Recently, chained BFT protocols~\cite{buchman2016tendermint, HotStuffYin2019, shi2019streamlined}, which periodically rotate leaders, were proposed for decentralized systems.
However, since the next leader is still predictable, the adversary can keep attacking the next leader to break the protocol's safety and/or liveness.
Single secret leader election protocols~\cite{boneh2020single, catalano2022adaptively} can be considered as mitigation, however, at the cost of extra overhead and further protocol complexity.

Apart from security properties, such leader-targeting attacks can trigger complex subprotocols for fail-over~\cite{buchman2016tendermint, casper, abraham2019sync} and view synchronization~\cite{HotStuffYin2019, shi2019streamlined} logic, leading to extra communication complexity and latency.
For example, PBFT's view change protocol~\cite{pbft1999} raises its communication complexity to $O(n^3)$, and requires multiple consecutive broadcasts to replace the previous leader.

\vspace{1em}
\noindent \textbf{BFT protocols are complex.}
Compared to Nakamoto-style consensus, BFT consensus design is notoriously complex~\cite{mickens2014saddest, NetComplex, underFire, 700BFT, Bessani2014}.
This complexity has directly affected its design, test, and deployment, which was well summarized by Guerraoui \etal~\cite{700BFT}: \textit{``They [BFT protocols] are notoriously difficult to develop, test and prove ... this difficulty, together with the impossibility of exhaustively testing distributed protocols~\cite{multipaxos} would rather plead for never changing a protocol that was tested and proven correct.''}
In contrast, blockchain developers often prefer to deploy protocols that are simple.
We believe that as a result of this, there are far fewer BFT-based blockchains~\cite{mapofcoins}.
As another example, in the context of Crash Fault Tolerant (CFT) consensus, the simpler Raft~\cite{raft2014} protocol is an increasingly popular choice~\cite{etcd, RedisRaft, CockroachDB} even though Paxos~\cite{lamport2001paxos} has been in use for many years.

The recent chained BFT protocols~\cite{buchman2016tendermint, HotStuffYin2019, shi2019streamlined} leverage the chain structure and pipelining to reduce the multi-phase voting process into a single-phase propose-vote scheme.
This is a significant simplification and leads to blockchain systems adopted in deployment~\cite{diem}.
However, the auxiliary fail-over~\cite{buchman2016tendermint, casper, abraham2019sync} and view synchronization~\cite{HotStuffYin2019, shi2019streamlined} subprotocols are proven to be bug-prone to design~\cite{cachin2017blockchain, ebb-and-flow, momose2019force} and challenging to implement~\cite{naor2019cogsworth, naor2020expected, bravo2020making}. This complexity continues in practice, hindering their adoption.

\begin{table}[t]
    \footnotesize
    \renewcommand{\arraystretch}{1.2}
    \caption{Comparison between \framework and existing BFT consensus protocols. Section~\ref{sec:related} provides a detailed analysis.}
    \begin{tabular}{r|cc|ccc}\hline
        \multicolumn{1}{l|}{}                & \makecell{Network                                  \\model}                           & Resil. & \makecell{Determ. \\safety}         & \makecell{Res. attacks\\on leader} & \makecell{No aux.\\ subprotocol}                    \\\hline
        Sync-HotStuff~\cite{abraham2019sync} & Sync.             & 1/2 & \cmark & \xmark & \xmark \\
        PBFT~\cite{pbft1999}                 & PSync.            & 1/3 & \cmark & \xmark & \xmark \\
        HotStuff~\cite{HotStuffYin2019}      & PSync.            & 1/3 & \cmark & \xmark & \xmark \\\hline
        Nakamoto~\cite{nakamoto2012bitcoin}  & Sync.             & 1/2 & \xmark & \cmark & \cmark \\
        GHOST~\cite{ghost}                   & Sync.             & 1/2 & \xmark & \cmark & \cmark \\\hline
        ByzCoin~\cite{ByzCoin}               & PSync.            & 1/3 & \cmark & \cmark & \xmark \\
        Pass and Shi~\cite{pass2017hybird}   & PSync.            & 1/3 & \cmark & \cmark & \xmark \\
        Ebb-and-flow~\cite{ebb-and-flow}     & PSync.            & 1/3 & \cmark & \cmark & \xmark \\\hline
        \textbf{\sysnameSyn}                 & Sync.             & 1/2 & \cmark & \cmark & \cmark \\
        \textbf{\sysnamePSyn}                & PSync.            & 1/3 & \cmark & \cmark & \cmark \\\hline
    \end{tabular}
    \label{table:comparison}
\end{table}

\vspace{1em}
\noindent \textbf{Nakamoto Consensus lacks deterministic safety.} The leader-based design of existing BFT protocols makes them vulnerable to targeted attacks, and also requires complex subprotocols to ensure security in the presence of Byzantine leaders.

Nakamoto-style consensus~\cite{nakamoto2012bitcoin} is an orthogonal approach to Byzantine consensus.
Contrary to traditional BFT protocols that only allow a leader to propose blocks, central to the design of Nakamoto-style consensus is \textit{egalitarianism}.
That is, any node can initialize a cryptographic lottery that commits to a certain predecessor block, and can produce a block after solving the lottery.
Nodes locally choose a fork (e.g., the longest fork) among the known forks to be the canonical chain.
The blockchains at different nodes converge on a block at time $t$ such that no other block is produced within $[t-\Delta, t+\Delta]$, where $\Delta$ is the network delay upper bound under synchronous networks.
Such a block is thus called the \emph{convergence opportunity}~\cite{niu2019analysis,UIUC}.

This egalitarian design makes the block proposer unpredictable and thus resistant to leader-targeting attacks; it also greatly simplifies the protocol.
Nakamoto-style consensus' fork choice rules allow convergence opportunities to happen regularly, ensuring safety.
However, a convergence opportunity is not predictable in Nakamoto-style consensus, making a node unable to finalize a block unless it becomes sufficiently deep in the blockchain.
This only ensures probabilistic safety, where the safety is less likely to be violated after a longer time.
This guarantee is strictly weaker than the deterministic safety achieved by BFT protocols and results in high latency~\cite{li2021close, gazi2022practical}, e.g., Bitcoin's famous one-hour confirmation rule~\cite{nakamoto2012bitcoin}.

\subsection{Our Proposal: Egalitarian BFT}
The above issues motivated us to design a consensus protocol for decentralized systems, e.g., blockchains, which gives the best of both worlds: 1) resistance to attacks on the leader, 2) simple design, and 3) deterministic safety and liveness guarantees,.
For the first two goals, we depart from the classic BFT leader-based design and follow the egalitarian approach inspired by Nakamoto-style consensus~\cite{nakamoto2012bitcoin} and EPaxos~\cite{Epaxos}, where any node can propose blocks.
For the last goal, we follow the design of classic BFT protocols by allowing a set of nodes (also known as a quorum) to vote for blocks.

\vspace{1em}
\noindent \textbf{EBFT: framework of egalitarian BFT protocols (\ssecref{sec:ebft}).}
We propose \framework, a framework for designing egalitarian BFT protocols with such guarantees.
Similar to Nakamoto-style consensus, all nodes continuously solve cryptographic lotteries (\eg, Proof-of-Work~\cite{nakamoto2012bitcoin}, Proof-of-Stake~\cite{Ouroboros}, verifiable delay functions~\cite{deb2021posat} and Proof-of-Elapsed-Time (PoET)~\cite{POET}), and any node solving a lottery can propose a block.
Unlike Nakamoto-style consensus where nodes are unaware of convergence opportunities, \framework allows nodes to proactively detect convergence opportunities, such that nodes can finalize a block associated to a convergence opportunity without the block being reverted later.
Detecting convergence opportunities allows \framework to achieve deterministic safety and liveness.
This egalitarian design also allows \framework to resist attacks on leaders, and rules out the need for any subprotocol to detect and replace faulty leaders.

We propose \sysnameSyn (\ssecref{subsec:syn}) and \sysnamePSyn (\ssecref{subsec:psyn}), two protocols in \framework equipped with mechanisms to detect convergence opportunities under synchronous and partially synchronous networks, respectively.
Both of them achieve the optimal resilience, i.e., 1/2 under synchronous networks and 1/3 under partially synchronous networks.
Moreover, due to the simplicity and strong resilience against leader-based attacks, one variant of \sysnamePSyn has been adopted in VeChain, an enterprise blockchain for supply chain management and business processes~\cite{vechain}.

\vspace{1em}
\noindent \textbf{\sysname with high performance (\ssecref{sec:ebft-opt}).}
In addition, we propose \sysname, a high-performance partially synchronous BFT protocol built on top of \sysnamePSyn.
Specifically, \sysname decouples transaction ordering from consensus by using techniques from Bitcoin-NG~\cite{bitcoinng}: once a node solves a cryptographic lottery, it proposes a fixed number of microblocks.
This improves the system's throughput and latency even while limiting the block production rate.
Table~\ref{table:comparison} provides a comparison between our three protocols and existing BFT consensus protocols.

\vspace{1em}
\noindent \textbf{Implementation and evaluation (\ssecref{sec:evaluation}).}
We implement these protocols based on the Go version of Bitcoin, \texttt{btcd}~\cite{btcd} and evaluated them on EC2.
We implemented \sysnameSyn, \sysnamePSyn and \sysname in 920 LoCs in total, which demonstrates their simple design.
Our evaluation results show that on a cluster of 256 geographically distributed nodes, \sysname achieves a throughput of 3.6k transactions per second and a latency of 8 seconds, which would satisfy the needs of many blockchain applications.

\section{Models and Preliminaries} \label{sec:model}
We consider a system of $n$ nodes that provides a Byzantine fault-tolerant service to a set of clients.
Each node has an index $i \in [n]$ where $[n] = \left \{1, 2, ..., n \right \}$, and the $i$-th identified node is denoted by $P_i$.
A subset of $f$ nodes may be Byzantine, \ie, behaving arbitrarily, at any time, whereas the remaining nodes are honest, \ie, strictly following the protocol.
There exists a public-key infrastructure (PKI) of nodes; each node $P_i$ has a pair of secret and public keys $(sk_i, pk_i)$ for signing and verifying messages.

\subsection{Threat Model}

\framework contains three protocols: \sysnameSyn, \sysnamePSyn and \sysname.
We assume that a minority of nodes (\ie, $f = \lfloor \frac{n}{2} \rfloor$) in \sysnameSyn are Byzantine,
and assume less than one third of nodes (\ie, $f = \lfloor \frac{n}{3} \rfloor$) in \sysnamePSyn and \sysname are Byzantine.
A probabilistic polynomial-time adversary controls these Byzantine nodes.
The adversary can get all Byzantine nodes' internal states and also can lead these nodes to arbitrarily misbehave during protocol execution.
The adversary can perform some probabilistic computing steps bounded by polynomials in the number of message bits generated by honest nodes.

We assume the adversary is \emph{adaptive}~\cite{pass2017hybird,pass2018thunderella, bentov2016snow, Ouroboros} in the sense that it can corrupt any set of $f$ nodes at any time.
Similar to existing blockchains with adaptive security~\cite{Ouroboros,pass2017hybird,pass2018thunderella, bentov2016snow}, we assume that honest nodes can implement erasures so that the adversary cannot access secret keys of nodes that used to be Byzantine.

We stress that \emph{adaptive security implies resilience against attacks on leaders.}
The adaptive adversary~\cite{Ouroboros, bentov2016snow, Ittai2019} is a well-established model for formalizing resistance against attacks based on adaptive corruption, including the attacks on a leader that is predictable.
Unlike a static adversary that only corrupts a set of $f$ nodes, the adaptive adversary can learn about the next leader, can corrupt this leader, and then directs the leader to launch more attacks.
Thus, if a consensus protocol is safe and live against an adaptive adversary, then it resists against attacks targeting the leader.

\subsection{Network Model}
We assume that there exist pairwise communication channels between every pair of nodes.
We consider two network models: synchronous for \sysnameSyn and partially synchronous for \sysnamePSyn and \sysname.

\begin{packeditemize}
    \item \textit{Synchronous network.}
    In this model, all messages between two nodes arrive within a given time bound $\Delta$.
    In other words, the entire execution is during a period of synchrony.

    \item \textit{Partially synchronous network.}
    In this model by Dwork \etal \cite{dwork1988consensus},
    there is a known delay bound $\Delta$ and an unknown Global Stabilization Time (\textsf{GST}) such that all message transmissions between two nodes arrive within the bound $\Delta$ after \textsf{GST}.
    In other words, the system is running in \textit{synchronous} mode after \textsf{GST} and in \textit{asynchronous} mode if \textsf{GST} never occurs. This model captures the impact of network partitions.
\end{packeditemize}

\subsection{Design Goals} \label{subsec:goals}

\vspace{1em} \noindent \textbf{Security properties.} 
Our goal is to design a simple and performant BFT consensus framework among $n$ nodes in the presence of $f$ static corruptions in synchronous or partially synchronous networks.
Specifically, the $n$ nodes receive transactions from clients and then commit client transactions into a totally ordered sequence.
The system provides the clients with an abstraction of a single non-faulty node and nodes only output non-duplicated transactions sent by clients.
Client transactions are packed into \textit{blocks}, and by committing a sequence of blocks, nodes can eventually observe the same sequence of transactions. These ordered blocks are eventually learned by the clients.
Formally, BFT consensus satisfies the following properties~\cite{abraham2019sync,HotStuffYin2019}:

\begin{packeditemize}
    \item \textit{Safety.} If any two honest nodes $P_1$ and $P_2$ output sequences of blocks $ \big <B_{0}, B_{1}, ..., B_{i} \big >$  and $\big < B_{0}^{\prime}, B_{1}^{\prime}, ..., B_{j}^{\prime} \big >$, respectively, then $B_{k} = B_{k}^{\prime}$ for $k \leq \min(i, j)$.

    \item \textit{Liveness.} Each client transaction will be eventually committed by all nodes.
\end{packeditemize}

We propose three protocols: \sysnameSyn that works in synchronous networks, and \sysnamePSyn and \sysname that work in partially synchronous networks.
\sysnameSyn guarantees safety and liveness in the synchrony model, while
\sysnamePSyn and \sysname guarantee safety  and guarantee liveness only after \textsf{GST}.

%
\vspace{1em}
\noindent \textbf{Performance Metrics.}
BFT consensus protocols concern two performance metrics: communication complexity and latency.

\begin{packeditemize}
    \item \textit{Communication complexity:} The total amount of data (in bits) transferred to commit a block.
    \item \textit{Latency:} The total amount of time taken to commit a block.
\end{packeditemize}


\subsection{Cryptographic Preliminaries}\label{subsec:preliminaries}
We assume standard cryptographic primitives are unbreakable, including hash functions and digital signatures.
A hash function $\textsf{H}(\cdot)$ takes an arbitrary-length string as its input and outputs a fixed-length bit string.
It has collision-resistance, \ie, finding two different messages $m_1$ and $m_2$ such that $\textsf{H}(m_1) = \textsf{H}(m_2)$ is computationally hard.
A signature scheme contains three functions:
the $\textrm{Gen}(1^\kappa)$ function takes an input of the security parameter $\kappa$ and outputs a public and private key pair $(pk, sk)$;
the $\textrm{Sign}(sk, msg)$ function outputs the signature $\sigma$ of the message $msg$ for a given private key $sk$;
the $\textrm{Verify}(pk, \sigma, msg)$ function takes a public key $pk$, signature $\sigma$, and message $msg$; it outputs $1$ if the signature is valid and $0$ if not.

We also assume cryptographic lottery, which takes a random string as its input for each round and outputs a proof $\tau$ if winning.
Meanwhile, there is a public algorithm that everyone can verify the winning proof.
Commonly used cryptographic lotteries include Proof-of-Work~\cite{nakamoto2012bitcoin}, Proof-of-Stake~\cite{Ouroboros}, verifiable delay functions~\cite{deb2021posat} and Proof-of-Elapsed-Time (PoET)~\cite{POET}.
A concrete implementation of cryptographic lotteries are provided in \ssecref{subsec:implementaion}.

\section{\framework Design}\label{sec:ebft}
In this section, we first present the design of \framework, including \sysnameSyn under synchrony (\ssecref{subsec:syn}) and \sysnamePSyn under partial synchrony (\ssecref{subsec:psyn}), and then provide formal security analysis (\ssecref{subsec:securityana}) and performance analysis (\ssecref{subsec:performanceana}).

\subsection{\sysnameSyn} \label{subsec:syn}

\sysnameSyn is a synchronous BFT protocol that works with the majority of nodes being honest.
We first provide an overview of \sysnameSyn, and then describe data structures and detailed protocol design.

\subsubsection{Overview} \label{subsec:syn-overview}
\begin{figure}[t]
    \centering
    \includegraphics[width=3in]{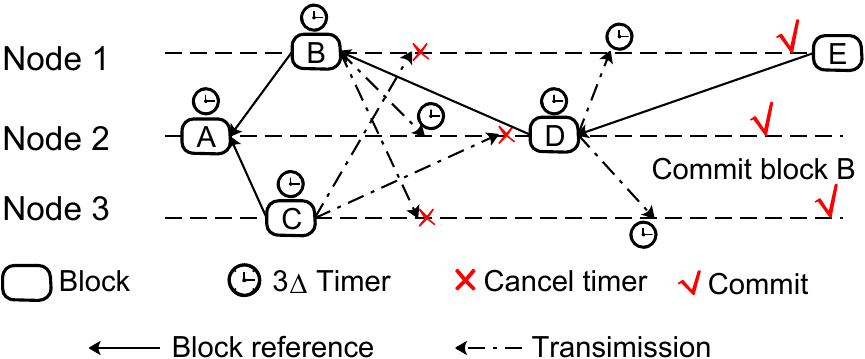}
    \setlength{\belowcaptionskip}{-0.5cm}
    \caption{\textbf{The consensus flow of \sysnameSyn.} Since blocks $C$ and $B$ conflict, nodes cancel the $3\Delta$ timer and do not commit any of them. Block $D$, together with uncommitted ancestor blocks, are committed by all nodes after the $3\Delta$ timer expires.
    }
    \label{fig:overview-syn}
\end{figure}

Figure~\ref{fig:overview-syn} provides an overview of the overall consensus flow in \sysnameSyn, in which nodes compete to produce blocks (\ie, egalitarian block production), follow the longest certified chain rule (LCCR) to vote and extend blocks, and use the $3\Delta$ timer to commit blocks. Next, we provide a high-level description of the above components to show the design intuition.

\noindent \textbf{Component \#1: Egalitarian block production.} Nodes in \sysnameSyn simultaneously participate in cryptographic lotteries to produce blocks. Once it wins a ticket, the node can produce a block that contains a set of transactions by following a certain format and then broadcasts the block together with the winning proof to others.

\vspace{1mm} \noindent \textbf{Component \#2: Longest certified chain rule.} When receiving a valid block, a node will vote for this block if it extends the longest certified chain in the node's local view. In \sysnameSyn, a block is certified if it is voted by enough nodes (\ie, at least $f+1$ nodes), and certified blocks are linked into a chain structure.
Meanwhile, nodes are asked to produce blocks after the longest certified chain.
The block voting and producing processes follow the same rule, which is referred to as the longest certified chain rule (LCCR). Note that if there are multiple longest certified chains, a node randomly chooses one to produce blocks.

The LCCR works in two aspects. First, following LCCR to produce blocks can resolve block conflict and eventually make all nodes agree on the longest one. Second, following LCCR to vote blocks guarantees that honest nodes refuse to vote for blocks not extending the longest one. This ensures that blocks that are already committed cannot be reverted due to a longer certified chain secretly created by the adversary.

\vspace{1mm} \noindent \textbf{Component \#3: $3\Delta$ committing Timer.}
When receiving a valid block that extends the longest certified chain in the local view, a node will set a timer of $3 \Delta$.
If the node does not receive any other blocks of the same height before the timer expires, then this is a convergence opportunity, and the node will commit the certified block together with all its uncommitted ancestor blocks.

Here we provide some intuition behind the $3 \Delta$ committing timer design.
In the synchrony model, a message sent at time $t$ by a node will be received by another node before time $t + \Delta$, where $\Delta$ is the maximum network delay.
The duration of $3 \Delta$ is to ensure that honest nodes can detect any conflicting blocks, by which they can commit certified blocks that are unique at their heights.
When a node observes a block that extends the longest certified chain at the time $t$, it then forwards the block to all other nodes.  Others will receive the block by time $t + \Delta$, and then honest nodes' votes will arrive at all nodes by time $t+ 2\Delta$.
Hence, the block will be certified by time $t+ 2\Delta$.
If any honest node votes for a conflicting block, it must do so before $t + 2\Delta$ (by LCCR).
As a result, nodes will receive the conflicting block by $t + 3\Delta$.

\subsubsection{Data Structure} \label{subsec:datastructure}
\noindent \textbf{Blocks and block format.} In \sysnameSyn, client transactions are batched into \textit{blocks}. In particular, blocks are linked into a chain structure, and each has the following structure:
\[B := \left( \text{Txs}, h_{p}, QC, \rho, meta \right),\]
where Txs is a collection of application-specific transactions; $h_{p}$ is the hash digest of the parent block; $QC$ is the parent block's quorum certificate; $\rho$ is the winning proof of the lottery; $meta$ represents necessary metadata.
There exists a hard-coded genesis block $\mathcal{G}_0 = \left(\text{Txs}, \perp, \perp, \perp, meta \right)$ and an associated $\mathcal{QC}_0$. Every block except $\mathcal{G}_0$ must specify its parent block by including the hash value and quorum certificate of that block. A block is \textit{valid} if it satisfies the following rules: $(i)$ it meets the block format; $(ii)$ its lottery proof is correct; $(iii)$ its parent block is valid; and $(iv)$ the including transactions meets the validity of applications and are not included in any previous blocks.

\vspace{1mm} \noindent \textbf{Vote and certificate.} A vote $v$ of block $B$ has the following structure:
\[v := \left ( h, pk, \sigma \right ),\]
where $h = \textsf{H}(B)$ is the hash of the block $B$; $pk$ is the node's public key; $\sigma$ is a signature created by the node over $\textsf{H}(B)$.
If there is a set of $f+1$ signatures on a block from nodes, then the block's quorum certificate (QC) is formed.
Here, a QC could be implemented as a simple set of individual signatures or aggregated signatures~\cite{aggregateSig}.
When a node has a QC for a block, the block is certified.
Each node keeps track of all signatures for all blocks and keeps updating certified blocks.

\vspace{1mm} \noindent \textbf{Block chaining and ranking.} Blocks are chained by a sequence of hash references and certificates. The chaining structure has been used in Bitcoin~\cite{nakamoto2012bitcoin} and state-of-the-art BFT protocols~\cite{buchman2016tendermint, HotStuffYin2019, shi2019streamlined}.
In particular, a block’s position (\ie the distance from the genesis block) in the chain is referred to as its height. The height of the genesis block is $0$.
A chain’s length is defined as the number of blocks in the chain excluding the genesis block.

A block $B$ is called a descendant of another block $B^{\prime}$ if $B$ extends a chain including $B^{\prime}$.
Conversely, block $B^{\prime}$ is an ancestor of block $B$. Two (distinct) blocks $B$ and $B^{\prime}$
conflict if neither is a descendant of the other.
Because of the possibility of conflicting blocks, each node maintains a block tree (referred to as $blockTree$) of received blocks.
In addition, certified blocks are ranked by their heights, and a certified block with the biggest height in the local $blockTree$ is referred to as the highest certified block.

\subsubsection{Protocol Description} \label{subsec:EBFT-Syn-desc}
Algorithm~\ref{alg:EBFT-Syn} illustrates the pseudocode of \sysnameSyn from a node's view.
It comprises four simple components: \textit{block producing}, \textit{block processing}, \textit{vote processing}, and \textit{timer interrupt processing}. These components can be realized by four event-driven functions and run in parallel.
These four components are outlined below.

\vspace{1mm} \noindent \textbf{Block producing.} Nodes participate in the cryptographic lottery to win the chance of producing blocks. Once won a ticket, the node can call the function \textsf{ProduceBlock}() (Line $14$-$20$).
In particular, the function first packages transactions into blocks and then includes the parent block's hash and QC. The parent block is the highest certified block that the node has seen.
The node first processes the block and then broadcasts it immediately.

\vspace{1mm} \noindent \textbf{Block processing.} When receiving a block, a node processes it by the function \textsf{ProcessBlock}().
In particular, if the block has already been stored, the function will return to avoid repetitive processing. Otherwise, the function will check the validity of the block, which includes verification of the block format, transactions, the parent block's hash and QC, and a nonce (see \ssecref{subsec:datastructure}).
If the block passes the validity check, the node will store this block and then call the function \textsf{ProcessCommitTimer}().
In this function, the node first checks if this block conflicts with other blocks at the same height.
If yes, the node cancels the timer of these conflicting blocks.
Otherwise, if this block extends the longest certified chain in local memory (\ie, the function \textsf{isSatisfyingLCCR}() returning true), the node will set a $3\Delta$ timer for it.
Besides, if this valid block also extends the longest certified chain, the node generates a vote for the block and then processes the vote.
After that, the node has to broadcast the block together with the associated vote. Here, the block is broadcast for honest nodes to detect conflicting blocks.
Note that a node may vote for multiple blocks at the same height if they all satisfy the above voting conditions.

\vspace{1mm} \noindent \textbf{Vote processing.} When receiving a vote, a node processes it by the function \textsf{ProcessVote}().
Specifically, if there is no valid block associated with the vote, the function will return.
Otherwise, it will check whether the vote has already been processed.
If not, the function will store the vote mapping with the block.
After that, if the associated block becomes certified, the node will update the highest certified block by the function $\textsf{UpdateHighestCertifiedBlock}()$.

\vspace{1mm} \noindent \textbf{Timer interrupt processing.} When a block's timer is triggered, the node commits this block together with all its non-committed ancestor blocks.

\begin{algorithm}[t]
    \small
    \caption{The pseudocode of \sysnameSyn protocol} \label{alg:EBFT-Syn}
    \begin{algorithmic}[1]
        \Statex \textbf{Local Variables}:
        \State {$M \leftarrow \{\mathcal{G}_0\}$  \textcolor{blue}{\Comment{the set of blocks}}}
        \State {$V \leftarrow \{ QC_0 \}$  \textcolor{blue}{\Comment{mapping votes with blocks}}}
        \State {$F \leftarrow \{\mathcal{G}_0 \}$  \textcolor{blue}{\Comment{the set of committed blocks}}}
        \State {$B^{\prime} \leftarrow \mathcal{G}_0$  \textcolor{blue}{\Comment{the highest certified block}}}
        \State $(pk, sk)$   \textcolor{blue}{\Comment{Key pair of a node}}
        \Statex
        \State
        \textbf{upon event} $\left \langle \texttt{Lottery-Win}|B \right \rangle$ \textbf{do}
        \State\hspace{\algorithmicindent}
        \textsf{ProduceBlock}() \textcolor{blue}{\Comment{producing blocks}}
        \Statex
        \State
        \textbf{upon event} $\left \langle \textsc{Block-Deliver}|B \right \rangle$ \textbf{do}
        \State \hspace{\algorithmicindent}
        \textsf{ProcessBlock}($B$) \textcolor{blue}{\Comment{processing receiving block}}
        \Statex
        \State
        \textbf{upon event} $\left \langle \textsc{Vote-Deliver}|v \right \rangle$ \textbf{do}
        \State \hspace{\algorithmicindent}
        \textsf{ProcessVote}($v$) \textcolor{blue}{\Comment{processing receiving vote}}
        \Statex
        \State
        \textbf{upon event} $\left \langle \textsc{Timer-Interrupt}|B \right \rangle$ \textbf{do}
        \State\hspace{\algorithmicindent}
        $F \leftarrow F \cup \textsf{GetAncestorBlocks}(B) \cup \{B\}$                     \textcolor{blue}{\Comment{committing blocks}}
        \Statex
        \State \textbf{procedure} $\textsf{ProduceBlock}()$
        \State\hspace{\algorithmicindent}
        $\textit{$B.Txs$} \leftarrow$ \textsf{GetTransactions}()
        \State\hspace{\algorithmicindent}
        $\textit{$B.h_p$} \leftarrow$ \textsf{H}($B^{\prime}$)                              \textcolor{blue}{\Comment{parent block's hash}}
        \State\hspace{\algorithmicindent}
        $\textit{$B.QC$} \leftarrow$ $V[B^{\prime}]$                                        \textcolor{blue}{\Comment{parent block's quorum certificate}}
        \State\hspace{\algorithmicindent}
        $\textit{$B.\rho$} \leftarrow \textsf{GetLotteryProof}()$
        \State \hspace{\algorithmicindent}
        \textsf{ProcessBlock}($B$)
        \State \hspace{\algorithmicindent}
        \textsf{Broadcast}($B$)
        \Statex
        \State \textbf{procedure} \textsf{ProcessBlock}($B$)
        \State\hspace{\algorithmicindent}
        \textbf{if} $\exists B \in M$  \textbf{then return}
        \State\hspace{\algorithmicindent}
        \textbf{if} \textsf{isValidNewBlock}($B$) \textbf{then}
        \State\hspace{\algorithmicindent} \hspace{\algorithmicindent}
        {$M \leftarrow M \cup \{B \}$}                                                     \textcolor{blue} {\Comment{already been stored}}
        \State \hspace{\algorithmicindent} \hspace{\algorithmicindent}
        \textsf{ProcessCommitTimer}($B$)
        \State \hspace{\algorithmicindent} \hspace{\algorithmicindent}
        \textbf{if} $\textsf{isSatisfyingLCCR}(B)$ \textbf{then}
        \hspace{\algorithmicindent}
        \State \hspace{\algorithmicindent}  \hspace{\algorithmicindent}
        \hspace{\algorithmicindent}
        $\sigma \leftarrow \textsf{Sig}(sk, \textsf{H}(B))$                                 \textcolor{blue}{\Comment{generating a signature}}
        \State \hspace{\algorithmicindent}  \hspace{\algorithmicindent}
        \hspace{\algorithmicindent}
        $v \leftarrow (\textsf{H}(B), pk, \sigma)$                                          \textcolor{blue}{\Comment{generating a vote}}
        \State\hspace{\algorithmicindent} \hspace{\algorithmicindent} \hspace{\algorithmicindent}
        \textsf{ProcessVote}($v$)
        \State\hspace{\algorithmicindent} \hspace{\algorithmicindent} \hspace{\algorithmicindent}
        \textsf{Broadcast}($B$, $v$)
        \Statex
        \State \textbf{procedure} \textsf{ProcessVote}($v$)
        \State\hspace{\algorithmicindent}
        \textbf{if} $\{ B | B \in M \wedge \textsf{H}(B)  = v.hash \} = \varnothing$ \textbf{then return}
        \State \hspace{\algorithmicindent}
        \textbf{if} $\exists v \in V[B]$  \textbf{then} \textbf{return}
        \State \hspace{\algorithmicindent}
        $V[B] \leftarrow V[B] \cup \{v\}$                                                   \textcolor{blue}{\Comment{storing the vote}}
        \State \hspace{\algorithmicindent}
        \textbf{if} $|V[B]| \geq f+1$ \textbf{then}
        \State \hspace{\algorithmicindent}  \hspace{\algorithmicindent}
        $B^{\prime} \leftarrow \textsf{UpdateHighestCertifiedBlock}()$
        \Statex
        \State \textbf{procedure} \textsf{ProcessCommitTimer}($B$)
        \State\hspace{\algorithmicindent}
        $S \leftarrow \{ B^{*} |  B \neq B^{*} \wedge B^{*}.height = B.height \}$           \textcolor{blue}{\Comment{conflicting blocks}}
        \State\hspace{\algorithmicindent}
        \textbf{if} $ S = \varnothing$ \textbf{then}
        \State \hspace{\algorithmicindent} \hspace{\algorithmicindent}
        \textbf{if} $\textsf{isSatisfyingLCCR}(B)$ \textbf{then}
        $\textsf{SetTimer}(B, 3\Delta)$
        \State \hspace{\algorithmicindent}
        \textbf{else then}
        \State \hspace{\algorithmicindent} \hspace{\algorithmicindent}
        \textbf{foreach} $B^{*} \in S$  \textbf{do}
        $\textsf{CancelTimer}(B^{*})$
    \end{algorithmic}
\end{algorithm}

\subsection{\sysnamePSyn} \label{subsec:psyn}
\sysnamePSyn is a protocol that extends \sysnameSyn to the partially synchronous network.
We first present an overview of required extensions, and then introduce chain structure and protocol design.

\subsubsection{Overview}
\begin{figure}[t]
    \centering
    \includegraphics[width=3in]{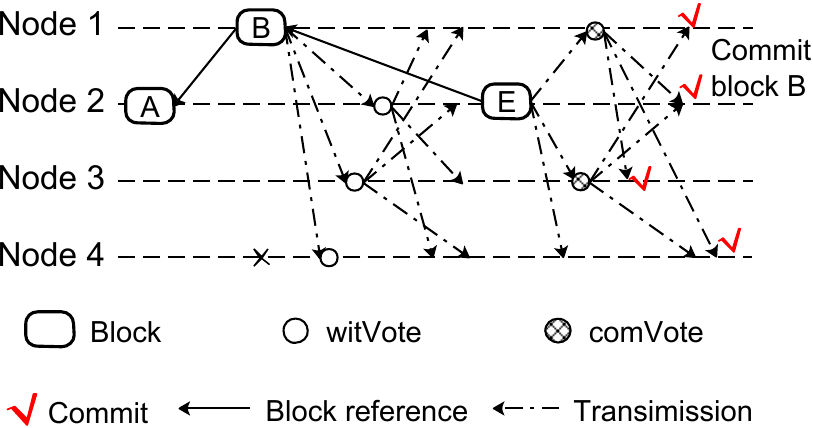}
    \setlength{\belowcaptionskip}{-0.5cm}
    \caption{\textbf{The consensus flow of \sysnamePSyn.} Nodes sends comVotes after receiving the block $E$ since the parent block $B$ is unique in its height. After receiving more than $2f+1$ comVotes, a node will commit the parent block $B$.
    }
    \label{fig:Psyn_flow}
\end{figure}

Figure~\ref{fig:Psyn_flow} provides an overview of the whole consensus flow in \sysnamePSyn.
\sysnamePSyn adopts the egalitarian block production and the longest certified chain rule (LCCR) in \sysnameSyn (see \textbf{Component \#1} and \textbf{Component \#2} in \ssecref{subsec:syn}).
Since there is no message delivery bound $\Delta$ before \textsf{GST}, \sysnamePSyn cannot rely on a timer to detect conflicting blocks.
Instead, \sysnamePSyn introduces another round of message exchanges for nodes to synchronize their views of non-conflicting blocks.
Due to the different network assumptions, a block is certified with at least $2f+1$ votes in \sysnamePSyn rather than $f+1$ in \sysnameSyn.
The remaining components in \sysnamePSyn are outlined below.

\vspace{1mm} \noindent \textbf{Component \#3: Committing blocks by uniqueness announcement.}
First, we introduce the term \textit{locally unique and certified} blocks.
In particular, if a block arrives at a node and becomes certified (\ie, enough votes being collected) before any other conflicting blocks at the same height are received by the node, the block is locally unique and certified. (The previously defined term \textit{uniquely certified blocks} refers to blocks that are globally unique and certified.)
When a node has a locally unique and certified block, it broadcast a uniqueness announcement of this block (\eg, another kind of vote).
The announcement denotes that the node will never cast vote for any other blocks at the same height if it follows the LCCR.
If a node receives at least $2f+1$ such announcements, then this is a convergence opportunity, and the node can commit the block and all its non-committed ancestor blocks.
Here, the threshold $2f+1$ guarantees that the majority of honest nodes have sent the announcements (since up to $f$ nodes are Byzantine), and so an adversary cannot collect $2f+1$ votes for any conflicting blocks to be certified at the same height.

\vspace{1mm} \noindent \textbf{Component \#4 (Optimization): Pipelining announcement and block voting.}
The uniqueness announcement of a certified block can be deferred to the voting processing of its child block (\ie, pipelining) such that we can make the protocol more efficient.
Therefore, there are two situations when voting for a block that satisfies LCCR.
If its parent block is uniquely certified, nodes can send votes that also carry the uniqueness announcement of its parent block.
Otherwise, its votes do not contain the uniqueness announcement.
To realize this, we differentiate vote types and introduce two kinds of votes: witness vote (\witVote for short) and commit vote (\comVote for short).
Meanwhile, we slightly revise the voting rule.
If receiving a new block extends the longest certified chain and the extended parent block is unique, a node casts a \comVote.
Otherwise, it casts a \witVote.
When a block is certified with at least $2f+1$ \comVotes, a node can commit all previous blocks of this block (except for this block).

Note that Component \#4 can make the voting process more efficient, at the cost of additional delay for committing blocks. Specifically, a block is committed when its child block is certified by including at least $2f+1$ \comVotes.

\subsubsection{Blockchain Structure}
\sysnamePSyn adopts the same block format and chain structure as these in \sysnameSyn. The slight difference lies in that there are two vote types and that a quorum certificate has to contain at least $2f+1$ distinct votes (rather than $f+1$ votes).
Specifically, a vote of the block $B$ has the following structure:
\[v := \left ( h, pk, type, \sigma \right ),\]
where an additional filed $type$ denote the type of the vote (\witVote or \comVote).
Here, the signature $\sigma$ is created by the node over $h$ and $type$.

\begin{algorithm}[t]
    \caption{The pseudocode of \sysnamePSyn protocol} \label{alg:EBFT-Syn+}
    \begin{algorithmic}[1]
        \Statex \textbf{Local Variables}:
        \State {$M \leftarrow \{\mathcal{G}_0\}$  \textcolor{blue}{\Comment{the set of blocks}}}
        \State {$V \leftarrow \{ QC_0 \}$  \textcolor{blue}{\Comment{mapping votes with blocks}}}
        \State {$F \leftarrow \{\mathcal{G}_0 \}$  \textcolor{blue}{\Comment{the set of committed blocks}}}
        \State {$B^{\prime} \leftarrow \mathcal{G}_0$  \textcolor{blue}{\Comment{the highest certified block}}}
        \State $(pk, sk)$   \textcolor{blue}{\Comment{Key pair of a node}}
        \Statex
        \State
        \textbf{upon event} $\left \langle \texttt{Lottery-Win}|B \right \rangle$ \textbf{do}
        \State\hspace{\algorithmicindent}
        \textsf{ProduceBlock}() \textcolor{blue}{\Comment{producing blocks}}
        \Statex
        \State
        \textbf{upon event} $\left \langle \textsc{Block-Deliver}|B \right \rangle$ \textbf{do}
        \State \hspace{\algorithmicindent}
        \textsf{ProcessBlock}($B$) \textcolor{blue}{\Comment{processing receiving block}}
        \Statex
        \State
        \textbf{upon event} $\left \langle \textsc{Vote-Deliver}|v \right \rangle$ \textbf{do}
        \State \hspace{\algorithmicindent}
        \textsf{ProcessVote}($v$) \textcolor{blue}{\Comment{processing receiving vote}}
        \Statex
        \State \textbf{procedure} $\textsf{ProduceBlock}()$
        \State\hspace{\algorithmicindent}
        $\textit{$B.Txs$} \leftarrow$ \textsf{GetTransactions}()
        \State\hspace{\algorithmicindent}
        $\textit{$B.h_p$} \leftarrow$ \textsf{H}($B^{\prime}$)                                 \textcolor{blue}{\Comment{parent block's hash}}
        \State\hspace{\algorithmicindent}
        $\textit{$B.QC$} \leftarrow$ $V[B^{\prime}]$                                            \textcolor{blue}{\Comment{parent block's quorum certificate}}
        \State\hspace{\algorithmicindent}
        $\textit{$B.\rho$} \leftarrow$ GetLotteryProof()
        \State \hspace{\algorithmicindent}
        \textsf{ProcessBlock}($B$)
        \State \hspace{\algorithmicindent}
        \textsf{Broadcast}($B$)
        \Statex
        \State \textbf{procedure} \textsf{ProcessBlock}($B$) \label{algo:processBlockPynStart}
        \State\hspace{\algorithmicindent}
        \textcolor{black}{ \textbf{if} $\exists B \in M$  \textbf{then return}}             \textcolor{blue}{\Comment{already been stored}}
        \State\hspace{\algorithmicindent}
        \textcolor{black}{\textbf{if} \textsf{isValidNewBlock}($B$) \textbf{then}}
        \State\hspace{\algorithmicindent} \hspace{\algorithmicindent}
        \textcolor{black}{$M \leftarrow M \cup \{B \}$}
        \State \hspace{\algorithmicindent} \hspace{\algorithmicindent}
        {\textbf{if} $\textsf{isSatisfyingLCCR}(B)$ \textbf{then}}
        \State \hspace{\algorithmicindent}  \hspace{\algorithmicindent}
        \hspace{\algorithmicindent}
        { \textbf{if} \textsf{isUniqueParentBlock}($B$) \textbf{then}}
        \State \hspace{\algorithmicindent}  \hspace{\algorithmicindent}
        \hspace{\algorithmicindent}
        \hspace{\algorithmicindent}
        {$\sigma \leftarrow \textsf{Sig}(sk, (\textsf{H}(B), \textsf{\comVote}))$}
            \State \hspace{\algorithmicindent}  \hspace{\algorithmicindent}
            \hspace{\algorithmicindent}
            \hspace{\algorithmicindent}
            {$v \leftarrow (\textsf{H}(B), pk, \textsf{\comVote}, \sigma)$ }
            \State \hspace{\algorithmicindent}  \hspace{\algorithmicindent}
            \hspace{\algorithmicindent}
            {\textbf{else} \textbf{then}}                                                    \textcolor{blue}{\Comment{generating a \textsf{\comVote}}}
            \State \hspace{\algorithmicindent}  \hspace{\algorithmicindent}
            \hspace{\algorithmicindent}
            \hspace{\algorithmicindent}
            { $\sigma \leftarrow \textsf{Sig}(sk, (\textsf{H}(B), \textsf{\witVote}))$}
        \State \hspace{\algorithmicindent}  \hspace{\algorithmicindent}
        \hspace{\algorithmicindent}
        \hspace{\algorithmicindent}
        {$v \leftarrow (\textsf{H}(B), pk,\textsf{\witVote}, \sigma)$}
        \State \hspace{\algorithmicindent} \hspace{\algorithmicindent}
        \hspace{\algorithmicindent}
        \textsf{ProcessVote}($v$)
        \State \hspace{\algorithmicindent} \hspace{\algorithmicindent} \label{algo:processBlockPynEnd}
        \hspace{\algorithmicindent}
        \textsf{Broadcast}($v$)
        \Statex
        \State \textbf{procedure} \textsf{ProcessVote}($v$)
        \State\hspace{\algorithmicindent}
        \textcolor{black}{\textbf{if} $\{ B | B \in M \wedge \textsf{H}(B)  = v.hash \} = \varnothing$ \textbf{then return} }
        \State \hspace{\algorithmicindent}
        \textcolor{black}{\textbf{if} $\exists v \in V[B]$  \textbf{then} \textbf{return}}
        \State \hspace{\algorithmicindent}
        \textcolor{black}{$V[B] \leftarrow V[B] \cup \{v\}$}                                 \textcolor{blue}{\Comment{storing the vote}}
        \State \hspace{\algorithmicindent}
        {\textbf{if} $|V[B]| \geq 2f+1$ \textbf{then}}
        \State \hspace{\algorithmicindent}  \hspace{\algorithmicindent}
        {$B^{\prime} \leftarrow \textsf{UpdateHighestCertifiedBlock}()$}
        \State \hspace{\algorithmicindent}
        {\textbf{if} $|V[B].\comVote| \geq 2f+1$ \textbf{then}}
        \State \hspace{\algorithmicindent}  \hspace{\algorithmicindent}
        {$F \leftarrow F \cup \textsf{GetAncestorBlock}(B)$}
    \end{algorithmic}
\end{algorithm}

\subsubsection{Protocol Design} \label{subsec:EBFT-Syn-desc+}
Algorithm~\ref{alg:EBFT-Syn+} illustrates the pseudocode of \sysnamePSyn, which comprises three key components: \textit{block producing}, \textit{block processing}, and \textit{vote processing}. Since block producing is the same as that in the synchronous network, we do not introduce it here. Please see \ssecref{subsec:EBFT-Syn-desc} for the detailed description.

\vspace{1mm} \noindent \textbf{Block processing.} When receiving a block, a node processes it by the function \textsf{ProcessBlock}() (Line 19-31).
The duplication and validity check is the same as that in Algorithm~\ref{alg:EBFT-Syn}.
The difference lies in the voting process for a block that satisfies the LCCR (Line $24-31$). In particular, if the node has voted for any other block at the same height as the block's parent block, it sends a \witVote and the hash of the previously voted block.\footnote{The included hash proofs can prevent Byzantine nodes from sending \witVotes on purpose. Removing this requirement does not affect committing blocks, since honest nodes will cast enough \comVotes.}
Otherwise, it sends a \comVote.
Note that for each block, a node only votes once, but a node can vote for multiple blocks at the same height if they all satisfy LCCR.

\vspace{1mm} \noindent \textbf{Vote processing.} When receiving a vote, a node processes it by the function \textsf{ProcessVote}() (Line $34$-$41$).
Specifically, if there is no valid block associated with the vote, the function will return.
Otherwise, it will check whether the vote has been processed.
If yes, it also returns.
If not processed previously, the function will store the vote and map the vote with the block.
After that, if the associated block becomes certified with no less than $2f+1$ (regardless of the vote types), the node will update the highest certified block by the function $\textsf{UpdateHighestCertifiedBlock}()$.
Besides, if a block has no less than $2f+1$ \comVotes, it will commit all non-committed ancestor blocks of this block, excluding this block.
Here, due to the pipelining structure, nodes actually commit the parent blocks of this block.

\subsection{Security Analysis} \label{subsec:securityana}
In this section, we provide a brief security analysis of \framework. Due to space constraints, we leave detailed proofs to Appendix~C and D. In particular, we prove that both \sysnameSyn and \sysnamePSyn satisfy \textit{safety} and \textit{liveness} properties;
the safety property guarantees that no two different blocks with the same height are committed, while the liveness property guarantees that clients' transactions will be eventually included in committed blocks no matter what the adversary does.

\subsubsection{\sysnameSyn} We only discuss the intuition behind the analysis here and leave the rigorous proof to Appendix~C.

\bheading{Safety}. In \framework, a block is committed directly or indirectly by its descendant node. Besides, a committed block must be certified.
In \sysnameSyn, if a block is directly committed by an honest node, the node must not receive any conflict block within $3\Delta$ and have collected $f+1$ votes for the committed block. By the strong $\Delta$-bounded assumption of the synchronous network, all honest nodes will receive the votes and certify the block before $2\Delta$. Therefore, no conflicting block will be voted by an honest node after that. If any node has voted for a conflicting block, it can only happen before $2\Delta$. Within $3\Delta$, all other nodes will observe the conflict block and cancel the committing timer.

\bheading{Liveness}. In \sysnameSyn, because the nodes controlled by the adversary are less than honest nodes, the block producing rate of the adversary is less than honest nodes. Therefore, we can show that with high probability, there always exists such uniquely certified blocks in a period $T$ no matter what the adversary does. As a result, the uniquely certified blocks will be committed, and its ancestor blocks will be indirectly committed.

\subsubsection{\sysnamePSyn} In \sysnamePSyn, there is no bounded delay for message delivery, so the safety and liveness analysis in \sysnamePSyn is slightly different from that in \sysnameSyn. The rigorous proof is provided in Appendix~D.

\bheading{Safety}. In \sysnamePSyn, an honest node casts \comVote for a block when its parent block is unique. The quorum size is $2f+1$ for committing a block. So there exists an honest node in the intersection of any two quorums. The parent block is certified. By the longest certified chain rule, the honest nodes do not vote for any block in conflict with its parent block. Therefore, once a block gets $2f+1$ \comVotes, any block in conflict with its parent block can not get $2f+$1 votes.

\bheading{Liveness.}
In \sysnamePSyn, when $\textsf{GST} = 0$, the case is the same as that in the synchronous network. When $\textsf{GST} > 0$, the adversary can withhold some blocks before \textsf{GST} due to the asynchronous network. We can show that the adversary can only withhold a finite number of blocks. Therefore, once the network is synchronous, by increasing the interval $T$, \sysnamePSyn can guarantee that there exist certified unique blocks, which will be committed with high probability.

\subsection{Performance Analysis} \label{subsec:performanceana}
In \sysnameSyn, \sysnamePSyn and \sysname, nodes have to broadcast newly receiving blocks to certify blocks.
This leads to a communication complexity of $O(n^2)$, which is the same as state-of-the-art leader-based BFT protocols like Dfinity~\cite{Hanke2018DFINITYTO}, Pili~\cite{Chan2019PiLiA} and Sync HotStuff~\cite{abraham2019sync}.

As shown in security analysis, the block interval can be set as $\Delta$ without affecting safety or liveness.
Thus, a transaction will be included in a new block within $0.5\Delta$ on average.
In \sysnameSyn, a block takes $3\Delta$ to be finalized, leading to the latency of $3.5\Delta$.
In \sysnamePSyn and \sysname, after \textsf{GST}, a block takes $2\Delta$ to be finalized, leading to the latency of $\textsf{GST} + 2.5\Delta$.

\begin{figure}[t]
    \centering
    \includegraphics[width=.4\textwidth]{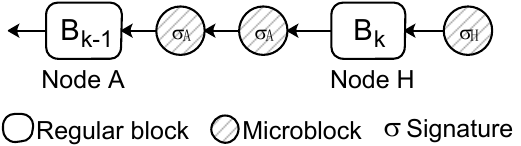}
    \setlength{\belowcaptionskip}{-0.5cm}
    \caption{\textbf{The blockchain structure of \sysname.} The creator of a regular block (denoted as a rectangle) can create a series of microblocks until the next regular block.}
    \label{fig:decouple}
\end{figure}

\section{\sysname}\label{sec:ebft-opt}
In this section, we introduce \sysname, which is built on \framework with optimized throughput.

\subsection{Overview}
In \framework, all nodes leverage the cryptographic lottery to win the rights to produce blocks. The probabilistic intrinsic of the lottery makes the interval between two consecutive blocks randomized, which inevitably causes forks, \ie, blocks sharing the same parent block.
Forking events will affect the commitment of blocks (see commit rule of \sysnameSyn and \sysnamePSyn in \ssecref{subsec:syn} and \ssecref{subsec:psyn}, respectively.), which further leads to a trade-off between latency and throughput.
That is, increasing the winning probability of the lottery (\ie, producing more blocks per second) will lead to a higher forking rate, and eventually, increases the commitment delay of blocks (and vice versa).
In other words, to keep the latency low, the average block interval will be large, limiting the system's throughput.

To address this issue, we leverage the observation that the network utilization is low during the empty period of two consecutive blocks. Thus, the block proposer could propose many consecutive blocks (called microblocks), which just contain some transactions, without going through the voting or committing rule. To distinguish them from microblocks, block are produced by winning the lottery and have to go through the certifying and committing rules (termed \textit{regular block}).
A simple illustration of these blocks is provided in Figure~\ref{fig:decouple}.
Similar ideas to differentiate blocks' functionalities have been used to improve system throughput~\cite{bitcoinng} and reward fairness~\cite{pass2017fruitchains, strongchain, Bissias2017BobtailAP} in NC.


\subsection{Protocol Design} \label{ebft-turbo:design}

\bheading{Blockchain Structure} \label{ebft-turbo:data}
In \sysname, there are two types of blocks: a regular block and a microblock.
Regular blocks are identical to blocks in \framework (see \ssecref{subsec:datastructure}), while mircoblocks have the following structure:
\[\textit{MicroB} := \left ( \text{Txs}, h, meta, \sigma \right),\] where $h$ is the hash of the previous block (either regular block or microblock), and $\sigma$ is the signature created by the node over all previous fields.
Vote messages are the same as in \framework.

\bheading{Algorithm description.}
Algorithm~\ref{alg:ebft-turbo} illustrates the modification to \sysnamePSyn's pseudocode required by \sysname. The main modifications are microblock production and processing functions, which are presented below.

\iheading{1) Microblock production.} After producing a regular block, a node can generate a series of microblocks at an allowed rate until the next regular block is produced.
In particular, once it produces a regular block, a node sets a timer to periodically produce microblocks (Lines 4-5 and 6-11).
Due to the introduction of microblocks, the proposing rule of regular blocks is slightly changed.
When producing regular blocks, nodes first choose the longest certified chain, which only considers certified regular blocks.
Then, nodes produce blocks after the latest microblock that extend the longest certified chain.
For example, in Figure~\ref{fig:decouple}, nodes first choose block $B_k$, and then produce regular blocks on the first microblock extending $B_k$.

\iheading{2) Microblock processing.}
When receiving a microblock, a node will check whether it is produced by the block owner of the highest certified regular block. If yes, it will store it and update the latest microblocks.

\iheading{3) Block committing.}
In \sysname, the committing rule for a regular block remains the same as in \framework.
Once a regular block is committed, all the ancestor regular blocks and microblocks are also committed.

All other mechanisms (\eg, regular block producing and processing) not listed in Algorithm~\ref{alg:ebft-turbo} are the same as in \framework.
\begin{algorithm}[t]
    \caption{The Pseudo-code of \sysname Protocol} \label{alg:ebft-turbo}
    \begin{algorithmic}[1]
        \State\textbf{upon event} $\left \langle \texttt{Lottery-Win}|B \right \rangle$ \textbf{do}
        \State\hspace{\algorithmicindent}
        \textsf{ProduceBlock}()
        \State\hspace{\algorithmicindent} $\textsf{SetMicroblockTimer}(B^{\prime}, v)$ \textcolor{blue}{\Comment{$v$: microblock interval}}
        \Statex
        \State
        \textbf{upon event} $\left \langle \textsc{Microblock-Timer-Interrupt}|B^{\prime} \right \rangle$ \textbf{do}
        \State\hspace{\algorithmicindent}
        \textbf{procedure} \textsf{ProduceMicroblock}()
        \Statex
        \State\textcolor{black}{\textbf{function} \textsf{ProduceMicroblock}($B^{\prime}$)}
        \State\hspace{\algorithmicindent}
        \textcolor{black}{$\textit{MircoB.Txs} \leftarrow$ \textsf{getTransactions}()}
        \State\hspace{\algorithmicindent}
        \textcolor{black}{$\textit{MircoB}.\sigma \leftarrow \textsf{Sig}(sk, \textsf{H}(MircoB.Txs))$}
        \State\hspace{\algorithmicindent}
        \textcolor{black}{$\textit{MircoB}.h \leftarrow \textsf{H}(B^{\prime})$} \textcolor{blue}{\Comment{$B^{\prime}$: the last block in chain}}
        \State \hspace{\algorithmicindent}  \textcolor{black}{\textsf{ProcessMicroblock}($MircoB$)}
        \State\hspace{\algorithmicindent} $\textsf{SetMicroblockTimer}(B^{\prime}, v)$ \textcolor{blue}{\Comment{update the timer}}
        \Statex

        \textcolor{black}{\State \textbf{function} \textsf{ProcessMicrBlock}($MircoB$)}
        \State\hspace{\algorithmicindent}
        \textcolor{black}{\textbf{if} $\exists \textit{MircoB} \in M$  \textbf{then return}}
        \State\hspace{\algorithmicindent}
        \textcolor{black}{\textbf{if} \textsf{isValidMicroBlock}($\textit{MircoB}$) \textbf{then}}
        \State\hspace{\algorithmicindent} \hspace{\algorithmicindent}
        \textcolor{black}{{$M \leftarrow M \cup \{B \}$}}
        \State \hspace{\algorithmicindent} \hspace{\algorithmicindent}
        \textcolor{black}{$B^{\prime} \leftarrow \textsf{updateHighestBlock}()$}
    \end{algorithmic}
\end{algorithm}

\subsection{Security Analysis}
In \sysname, the introduction of microblocks do not affect the committing rule of \framework (including \sysnameSyn and \sysnamePSyn). Thus, \sysname satisfies the same safety and liveness properties as \framework.

\section{Implementation and Evaluation} \label{sec:evaluation}
To demonstrate the simplicity and practicality of EBFT, we implement \framework and \sysname, and then evaluate their performance on a cluster of Amazon EC2 instances.
We conduct two groups of experiments:
one is on a small cluster of 16 instances for making comparisons with HotStuff~\cite{HotStuffYin2019},
and the other is on a large cluster of up to 256 instances for demonstrating the practicality in large-scale deployments.
The former group of experiments shows that \framework achieves about half the latency of HotStuff under the same throughput of >1000 transactions per second.
The latter group of experiments shows that running across 256 instances, \sysname processes 3200 transactions per second and commits a transaction in 8 seconds.

Our evaluation aims at answering the following questions:
\begin{packeditemize}
    \item \textbf{Simplicity:} How much effort, quantified in lines of code (LoCs), is needed to implement \framework protocols?
    \item \textbf{Throughput/latency v.s. HotStuff:} How do \framework protocols compare with the state-of-the-art HotStuff consensus~\cite{HotStuffYin2019} in terms of throughput and latency, in the best case and under attacks?
    \item \textbf{Throughput/latency at scale:} What are the maximum throughput and latency that \framework and \sysname can achieve under a large-scale deployment?
\end{packeditemize}

In addition, we are also interested in the following empirical metrics, which provide insight into the performance of our protocols.

\begin{packeditemize}
    \item \textbf{Block propagation delay} is the time needed for a newly produced block to be propagated to the entire network.
    \item \textbf{Network utilization} is the utilized bandwidth during the protocol execution.
    \item \textbf{Forking rate} is the ratio of the number of committed blocks over the number of total produced blocks.
\end{packeditemize}

\subsection{Implementation} \label{subsec:implementaion}
We have implemented two variants of \framework.
First, we provide an implementation of \sysnamePSyn based on the \texttt{bamboo} prototyping framework~\cite{chainedBFT}, in order to fairly compare with HotStuff~\cite{HotStuffYin2019} on the same platform.
Second, we provide an implementation of \sysnameSyn, \sysnamePSyn and \sysname based on \texttt{btcd}~\cite{btcd}, a production-level Bitcoin implementation in Go, in order to demonstrate the simplicity and practicality of our protocols.
On top of \texttt{btcd}, \sysnameSyn, \sysnamePSyn and \sysname take about 600, 120 and 200 extra LoCs, respectively, leading to 920 LoCs added/modified in total.
Both of our implementations are open source~\cite{impl, impl-bamboo}.

\vspace{1mm}
\noindent \textbf{Implementation based on \texttt{bamboo}.}
The \texttt{bamboo} project~\cite{chainedBFT} is a framework for prototyping chained BFT protocols in Go.
It provides programming interfaces for four components in chained BFT protocols: block proposal, voting rule and commit rule.
We implement \sysnamePSyn by these interfaces.
Specifically, for block proposal, we set each node to have the same block producing rate.
For voting rule, each node votes for the longest certified chain in its view.
For commit rule, each node finalizes a certified block when it receives enough uniqueness announcement votes.

\vspace{1mm}
\noindent \textbf{Implementation over \texttt{btcd}.}
The \texttt{btcd} project is a production-level implementation of Bitcoin in Golang.
We implement \sysnameSyn, \sysnamePSyn and \sysname on top of \texttt{btcd} release 0.22.0.
One notable difference with the \texttt{bamboo}-based implementation is the peer-to-peer network.
While \texttt{bamboo} enforces a fully connected network, \texttt{btcd} allows nodes to propagate messages through a peer-to-Peer network, in which a node can only be directly connected to a small subset of peers.
In \texttt{btcd}, thus our implementation, a node by default has at most 8 outbound connections.
Consequently, some nodes at the edge of the network may not be able to receive broadcast messages.
In our protocols, if such nodes cannot collect enough votes for a block, then these nodes will lose liveness.
To make votes to be received by as many nodes as possible, the implementation requires nodes to proactively forward received votes to their peers.

We implement \sysnameSyn in about 600 LoCs, \sysnamePSyn in about 120 extra LoCs, and \sysname in about 200 extra LoCs, leading to 920 LoCs added/modified in total, over \texttt{btcd}.
Table~\ref{table:locs-changed-added} provides a summary of detailed changes compared to \texttt{btcd}.
This demonstrates the simplicity of our protocols for implementing on production-level blockchain platforms.
Note that The server.go file contains all functionalities for peer communication. For better code clarity and readability, we create the serverpeer.go file and remove necessary functionalities that are originally in the server.go file to this file.

\begin{table}[t]
    \centering
    \caption{Summary of LoCs deleted/added compared to \texttt{btcd} release 0.22.0~\cite{btcd}.}
    \begin{threeparttable}
        \small
        \begin{tabular}{@{}l c | l c@{}}
            \toprule[1pt]
            File                     & \makecell{Del./Add.                                  \\ Locs}  & File & \makecell{Del./Add. \\ Locs}\\
            \midrule
            blockchain/accept.go     & 5/5                 & config.go            & 2/66    \\
            blockchain/blockindex.go & 10/39               & limits\_plan9.go     & 10/0    \\
            blockchain/chain.go      & 2/32                & limits\_unix.go      & 52/0    \\
            blockchain/chainio.go    & 0/21                & netsync/interface.go & 0/5     \\
            blockchain/committee.go  & 0/82                & peer/peer.go         & 1/9     \\
            blockchain/orazor.go     & 0/231               & rpcserver.go         & 0/66    \\
            blockchain/process.go    & 1 /50               & server.go            & 1176/54 \\
            blockchain/weight.go     & 2/2                 & wire/common.go       & 0/ 31   \\
            chaincfg/extension.go    & 0/102               & wire/message.go      & 0/ 4    \\
            chaincfg/params.go       & 0/5                 & wire/msgvote.go      & 0/69    \\
            limits\_windows.go       & 10/0                & serverpeer.go        & 0/1164  \\
            netsync/manager.go       & 14/164              & wire/msgblock.go     & 1/2     \\
            \bottomrule[0.9pt]
        \end{tabular}
    \end{threeparttable}
    \label{table:locs-changed-added}
\end{table}

\vspace{1mm}
\noindent \textbf{Cryptographic lottery}. In \sysname, nodes participate in a cryptographic lottery to win the rights to produce blocks.
The cryptographic lottery is widely used in Nakamoto-style blockchains to control the block production rate~\cite{ebb-and-flow, nakamotoRace}.
The implementation of the cryptographic lottery can be further divided into cryptography-based solutions and secure-hardware-based solutions.
The widely used cryptography-based solutions include Proof-of-Work (PoW)~\cite{nakamoto2012bitcoin, garay2018blockchain}, Proof-of-Stake~\cite{Ouroboros}, verifiable delay function (VDF)~\cite{deb2021posat}, and Proof-of-Space (PoSpace)~\cite{Hamza2017, Sunoo2018}.
The secure-hardware-based solution is Proof-of-Elapsed-Time (PoET)~\cite{POET, wangmulti}.
The \texttt{btcd} codebase natively supports the proof-of-work (PoW)-based lottery in Bitcoin.
It also provides interfaces for simulating the block production process without actual mining, via the command line \texttt{btcctl generate}.
During our experiments, we set each node to have the same block producing rate.

\subsection{Experimental Setup}

We evaluate the performance of these protocols on Amazon's EC2 instances.
Specifically, we deploy our protocols over 256 \texttt{t2.micro} instances (1 GB RAM, one CPU core, and 60-80 Mbit/s network bandwidth) in 13 regions around the globe\footnote{The regions include North Virginia, North California, Oregon, Ohio, Canada, Mumbai, Seoul, Sydney, Tokyo, Singapore, Ireland, Sao Paulo, London, and Frankfurt.}.
Each instance hosts a single node, as \texttt{btcd} provides little support for multiplexing on the same computer.
Due to CPU constraints imposed by AWS, our implementation does not verify transactions, but instead fills each transaction with 512 random Bytes.
In addition, the implementation does not employ aggregation techniques for signature signing/validation.
We use a fixed committee in \sysnameSyn and \sysnamePSyn.
We consider three committee sizes of 64, 128, and 256, four block sizes of 20, 40, 80, 160 \kbytes, and the average block interval of 2 seconds.

\subsection{Throughput/Latency vs. HotStuff}
\label{subsec:eval-bamboo}

We first evaluate the throughput and latency of \sysnamePSyn and HotStuff~\cite{HotStuffYin2019} by using the \texttt{bamboo} framework.
We use the block interval of $5$s, the block size of 400\kbytes, and the microblock rates of 4 blocks per epoch over nodes, and deploy the system on 16 AWS EC2 instances.

\begin{figure}[t]
    \centering
    \includegraphics[width=.85\linewidth]{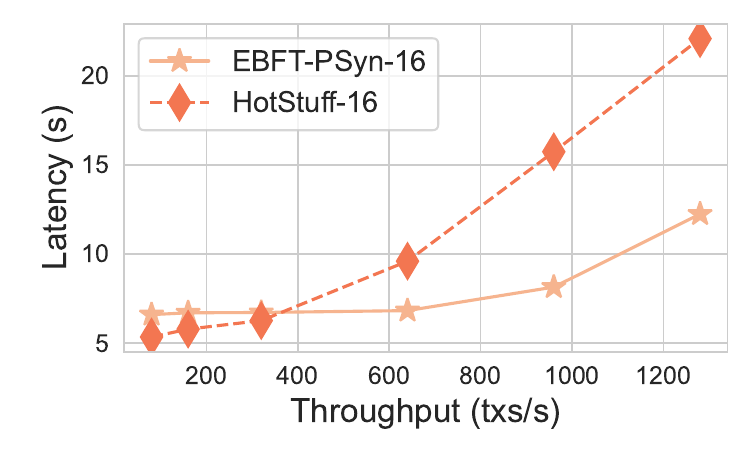}
    \caption{\sysnamePSyn v.s. HotStuff with no Byzantine node.}
    \label{fig:vs-hs}
\end{figure}

Figure~\ref{fig:vs-hs} shows the throughput and latency of \sysnamePSyn and HotStuff when all nodes are honest.
The evaluation results show that \sysnamePSyn achieves comparable throughput and latency with HotStuff.
Specifically, when the throughput is less than 6400 transactions per second, HotStuff achieves better latency, but after that HotStuff's latency increases significantly compared to \sysnamePSyn.
This is because HotStuff is responsive while having a higher concrete communication overhead than \sysnamePSyn.
When the throughput is small and the bandwidth is not fully utilized, the real-time network latency is small, so HotStuff achieves a smaller latency.
Meanwhile, \sysnamePSyn is not responsive and produces a block every $100$ms.
When the throughput is large and the bandwidth is almost fully utilized, the real-time network latency becomes larger, so that HotStuff's latency increases significantly.
Since \sysnamePSyn has less concrete communication overhead, its latency then becomes superior to HotStuff.

\begin{figure}[t]
    \centering
    \begin{subfigure}[]{0.48\linewidth}
        \centering
        \includegraphics[width=\linewidth]{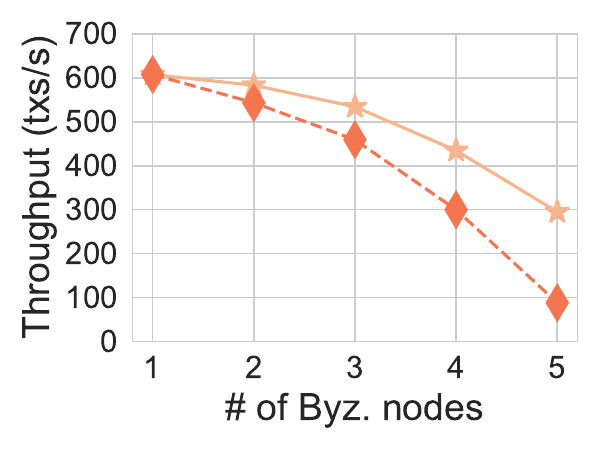}
        \caption{Throughput under forking attack.}
        \label{fig:f-tps-vs-hs-fork}
    \end{subfigure}
    \hfill
    \begin{subfigure}[]{0.48\linewidth}
        \centering
        \includegraphics[width=\linewidth]{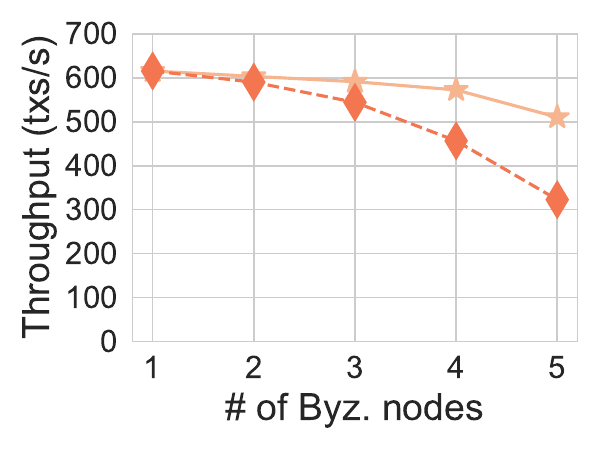}
        \caption{Throughput under silence attack.}
        \label{fig:f-tps-vs-hs-silence}
    \end{subfigure}

    \begin{subfigure}[]{0.48\linewidth}
        \centering
        \includegraphics[width=\linewidth]{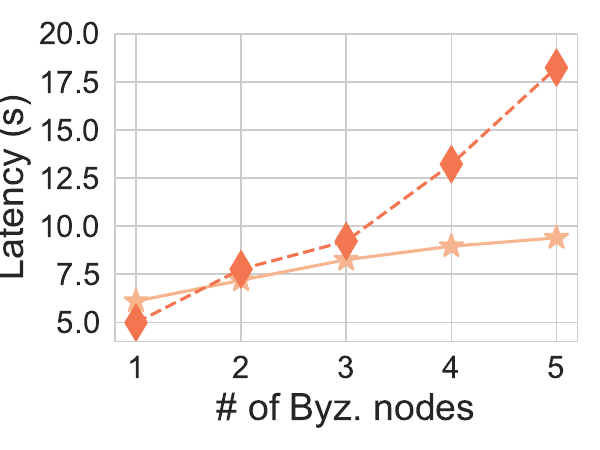}
        \caption{Latency under forking attack.}
        \label{fig:f-latency-vs-hs-fork}
    \end{subfigure}
    \hfill
    \begin{subfigure}[]{0.48\linewidth}
        \centering
        \includegraphics[width=\linewidth]{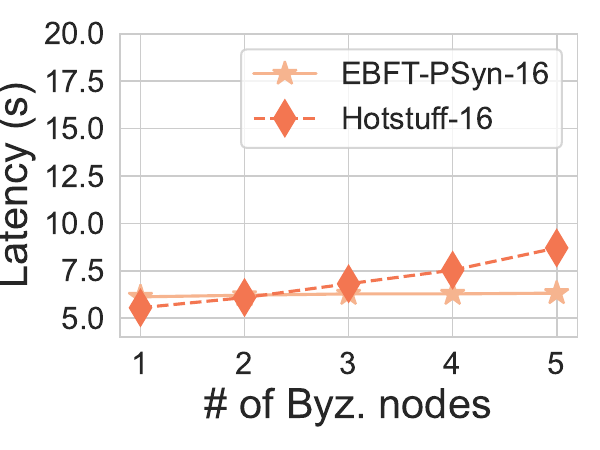}
        \caption{Latency under silence attack.}
        \label{fig:f-latency-vs-hs-silence}
    \end{subfigure}
    \caption{\sysnamePSyn vs Hotstuff under attacks.}
    \label{fig:ebft-vs-hs-attacks}
\end{figure}

Figure~\ref{fig:ebft-vs-hs-attacks} shows the throughput and latency of \sysnamePSyn and HotStuff under attacks launched by up to 5 nodes.
We simulated two types of attacks:
\emph{forking attack} where Byzantine nodes propose conflicting blocks, and
\emph{silence attack} where Byzantine nodes stop sending any message\cite{chainedBFT, niu2021performance}.
In terms of throughput, the results show that \sysnamePSyn achieves better throughput than HotStuff under both forking and silence attacks.
This is because \sysnamePSyn commits a block within two broadcast rounds, which gives the adversary less opportunity to overwrite a block or delay the formation of quorums.
This is consistent with observations in the \texttt{bamboo} paper~\cite{chainedBFT}, where two-chain rules are more resilient against attacks than three-chain rules.
In terms of latency, the results show that \sysnamePSyn achieves worse latency than HotStuff when $f$ is $1\sim2$, but achieves better latency when $f$ becomes larger than $2$.
When $f$ is small, HotStuff commits blocks faster than \sysnamePSyn since HotStuff is responsive, i.e., commits blocks at real-time latency.
When $f$ becomes larger, as \sysnamePSyn is more resilient to attacks, attacks will have a lower impact on latency in \sysnamePSyn as compared to HotStuff.

\subsection{Large-scale Experiments}

We then evaluate the throughput and latency of \sysnameSyn, \sysnamePSyn and \sysname in a large-scale deployment by using the \texttt{btcd}-based implementation.
In addition, we evaluate the block propagation delay, forking rate, network utilization and latency under different block sizes.

\begin{figure}[t]
    \centering
    \includegraphics[width=.6\linewidth]{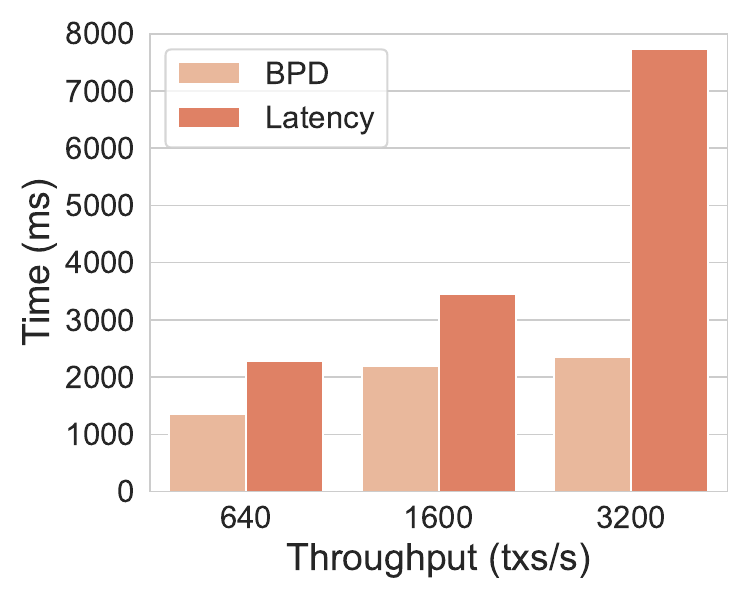}
    \caption{Throughput vs. latency of \sysname with 256 nodes.}
    \label{fig:tps}
\end{figure}

\vspace{1mm}
\noindent \textbf{Throughput and latency.}
Figure~\ref{fig:tps} shows the throughput and latency of \sysname under the regular block interval of $2$\secs, the block size of 160\kbytes, and the microblock rates of \{4, 10, 20\} blocks per epoch.
Given that each transaction takes 512 Bytes and the block size of 160\kbytes, these microblock rates lead to \{640, 1600, 3200\} transactions per second.
Note that \sysname follows Bitcoin's P2P network unlike in \ssecref{subsec:eval-bamboo}.
The results show that increasing the rate of microblocks leads to a higher throughput, but at the cost of slightly increased block propagation delay and latency.
These show that \sysname can achieve both high throughput and low latency.

\begin{figure}[t]
    \centering
    \includegraphics[width=.85\linewidth]{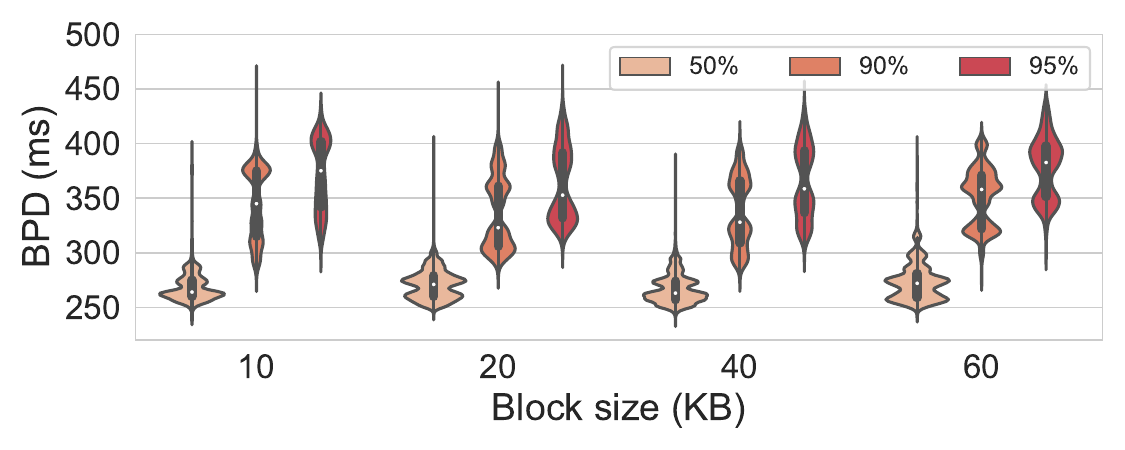}
    \caption{Block propagation delay.}
    \label{fig:bpd}
\end{figure}

\vspace{1mm}
\noindent \textbf{Block propagation delay.}
Figure~\ref{fig:bpd} shows the time distribution for blocks to propagate to $50\%$, $90\%$ and $95\%$ of nodes for blocks of different sizes. We set the block interval to $2$\secs.
The $50\%$ block propagation latency is concentrated at 250-300ms.
The $90\%$ and $95\%$ block propagation latency is within 500 ms.
This shows that increasing block sizes can slightly affect the block propagation delay.

\begin{figure}[t]
    \centering
    \includegraphics[width=.7\linewidth]{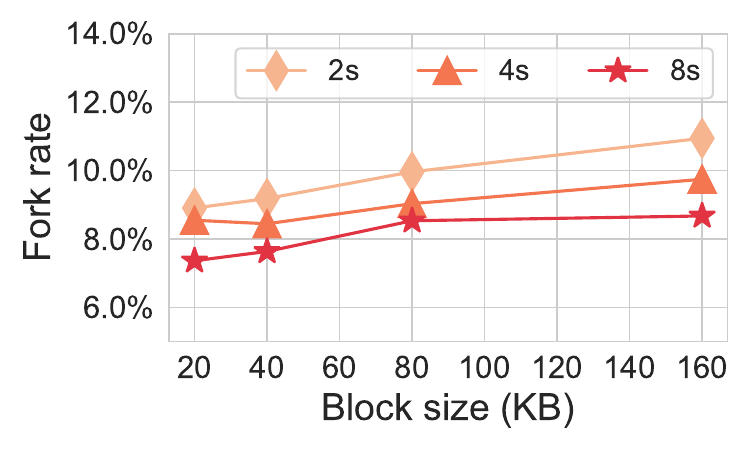}
    \caption{Forking rate of \sysnameSyn.}
    \label{fig:fork-rate}
\end{figure}

\vspace{1mm}
\noindent \textbf{Forking rate.}
Figure~\ref{fig:fork-rate} displays the forking rate of \sysnamePSyn under a committee of 256 nodes with the block sizes of \{20, 40, 80, 160\} Bytes and the block intervals of $\{ 2, 4, 8\}$ \secs.
The forking rate is quantified by the ratio between the number of blocks that are not in the committed chain and the number of all produced blocks.
We observe that both increasing the block size and reducing the block interval result in a higher forking rate.
As the forking rate remains less than 12\% even with a block interval of $2$\secs, we set the block interval to $2$\secs in our subsequent experiments.

\begin{figure}[t]
    \centering
    \begin{subfigure}[]{0.48\linewidth}
        \centering
        \includegraphics[width=\linewidth]{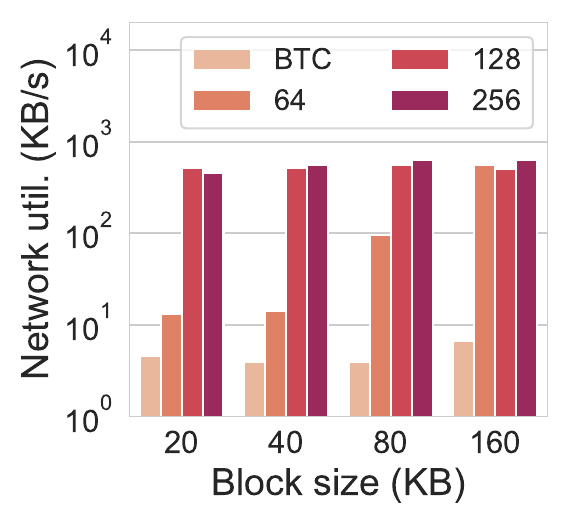}
        \caption{\sysnameSyn.}
        \label{fig:bw-util-syncorazor}
    \end{subfigure}
    \hfill
    \begin{subfigure}[]{0.48\linewidth}
        \centering
        \includegraphics[width=\linewidth]{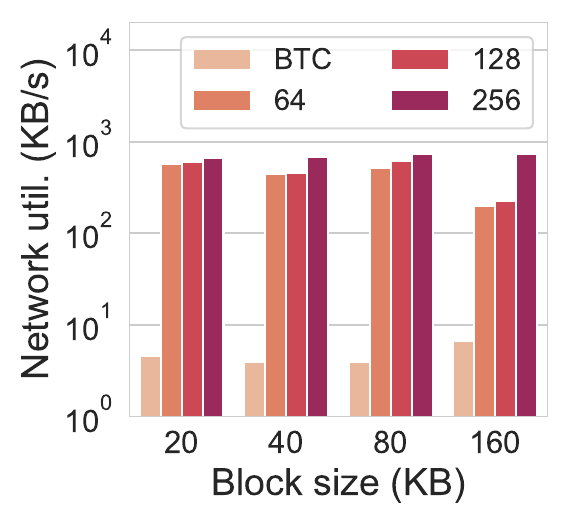}
        \caption{\sysnamePSyn.}
        \label{fig:bw-util-psyncorazor}
    \end{subfigure}
    \caption{Nodes' network utilization.}
    \label{fig:bw-util}
\end{figure}

\vspace{1mm}
\noindent \textbf{Network utilization.}
Figure~\ref{fig:bw-util-syncorazor} shows the network utilization with the block interval of $2$\secs, the block sizes of $\{20, 40, 80, 160\}$\kbytes, and the committee sizes of $\{ 64, 128, 256 \}$, with a comparison to a cluster of $256$ Bitcoin nodes running \texttt{btcd}.
While each node in BTC utilizes about $6$\kbytes/s, each node in \sysnameSyn and \sysnamePSyn utilize a constant bandwidth of $\approx 600$\kbytes/s per second, except that \sysnameSyn with the committee size of $64$ utilizes more bandwidth with increasing block sizes.
This is because the major overhead is propagating blocks in this setting.

\begin{figure}[t]
    \centering
    \begin{subfigure}[]{0.48\linewidth}
        \centering
        \includegraphics[width=1.1\linewidth]{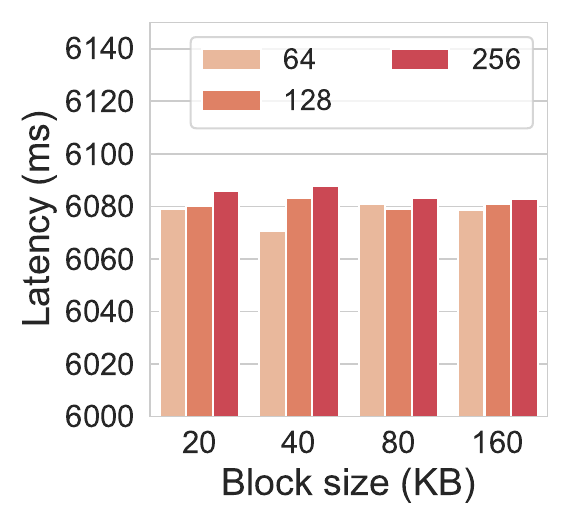}
        \caption{\sysnameSyn.}
        \label{fig:latency-syncorazor}
    \end{subfigure}
    \hfill
    \begin{subfigure}[]{0.48\linewidth}
        \centering
        \includegraphics[width=1.1\linewidth]{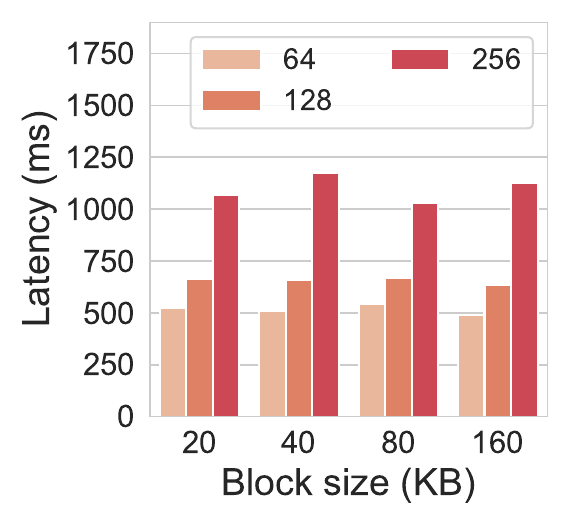}
        \caption{\sysnamePSyn.}
        \label{fig:latency-psyncorazor}
    \end{subfigure}
    \caption{Latency.}
    \label{fig:latency}
\end{figure}

\vspace{1mm}
\noindent \textbf{Latency vs. block size.}
Figure~\ref{fig:latency} plots the average latency, \ie the time taken for a block from being produced to being committed, under the block interval of $2$\secs.
These results show that a block is committed in less than $6.1$\secs and $1.2$\secs in \sysnameSyn and \sysnamePSyn (without using pipelining), respectively.
By contrast, the latency of Nakamoto-style blockchains has to take tens of block intervals to confirm a transaction~\cite{gervais2016security}.
For example, in Ethereum, blocks are produced every 10-20\secs, and nodes have to wait for 12 produced blocks, which leads to more than 4 minutes.

\section{Related Work} \label{sec:related}
In this section, we discuss existing protocols, including BFT protocols, Nakamoto-style consensus protocols, and hybrid consensus that combine BFT and Nakamoto-style consensus.
Table~\ref{table:comparison} shows that \framework is the only family of consensus protocols that achieve deterministic safety and resilience against attacks on leaders, without the need for an auxiliary sub-protocol.

\subsection{BFT Protocols}

Castro and Liskov~\cite{pbft1999} proposed PBFT, the first practical leader-driven BFT consensus protocol.
The design of PBFT has inspired a batch of leader-based BFT protocols~\cite{ kotla2007zyzzyva, HQReplica, 700BFT}.
However, as analyzed in~\ssecref{sec:introduction}, the leader-based design makes these protocols still vulnerable to attacks targeted at the leader, and requires complex subprotocols such as state synchronization and fail-over protocols for detecting and replacing the Byzantine leader.
On contrary, \framework does not need these complicated subprotocols and is more resilient to attacks targeted at the leader.

Recently, many synchronous BFT protocols~\cite{Hanke2018DFINITYTO, Chan2019PiLiA, abraham2019sync} and partially synchronous BFT protocols~\cite{ buchman2016tendermint, HotStuffYin2019, abraham2019sync, chan2018pala, shi2019streamlined} in the arena of blockchains have been proposed.
However, since the next leader is still predictable, these protocols are still subjected to attacks on the leader, and require subprotocols for replacing the Byzantine leader.
Single secret leader election~\cite{boneh2020single} protocols can be considered as a mitigation~\cite{catalano2022adaptively}, however at the cost of extra overhead and further protocol complexity.

\subsection{Nakamoto-style Consensus}
Nakamoto-style consensus, first proposed in Bitcoin~\cite{nakamoto2012bitcoin}, is an orthogonal approach to Byzantine consensus.
Contrary to traditional BFT protocols that only allow a leader to propose blocks, Nakamoto-style consensus allows any node to initialize a cryptographic lottery that commits to a predecessor block, and can produce a block after solving the lottery.
Nodes locally choose a fork (e.g., the longest fork) among the known forks to be the canonical chain.
Following Bitcoin, a number of Nakamoto-style consensus protocols with different trade-offs~\cite{ghost, Sompolinsky2016SPECTREAF, sompolinsky2018phantom} have been proposed.

However, Nakamoto-style consensus protocols are proven~\cite[Theorem 5.1]{lewis2021CCS} to only achieve probabilistic safety guarantee, where the probability that a block is reverted decreases exponentially with its depth~\cite{UIUC, niu2019analysis, nakamotoRace}.
The probabilistic safety guarantee is strictly weaker than the deterministic one achieved in BFT protocols, where a committed block can never be reverted.

\subsection{Hybrid Consensus}
There have been proposals to combine BFT and Nakamoto-style consensus to get the best of both worlds.
Byzcoin~\cite{ByzCoin} is among the first to incorporate BFT protocol with Nakamoto-style consensus to achieve deterministic safety in a synchronous network.
In Byzcoin, nodes who have produced a history of the last blocks in the chain form the committee and run PBFT~\cite{castro1999practical} to commit transactions.
However, due to forks, the selected committee may not be committed, which further compromises the finality guarantee of transactions~\cite{pass2017hybird}.
Pass and Shi~\cite{pass2017hybird} address this issue by using a fragment of committed blocks (confirmed by a large $k$) to construct the committee.
Later, Buterin and Griffith~\cite{casper} propose Casper FFG, which utilizes a pipelined BFT protocol as the finality layer and extends Nakamoto-style consensus to the partially synchronous model.
Inspired by Casper FFG, several protocols like Afgjort~\cite{dinsdale2020afgjort}, GRANDPA~\cite{stewart2020grandpa}, Snap-and-Chat protocols~\cite{ebb-and-flow} as well as the Checkpointed Longest Chain~\cite{sankagiri2020blockchain} are proposed to combine off-the-shelf leader-based BFT protocol with Nakamoto-style consensus.

These protocols achieve safety and liveness under partially synchronous networks.
However, despite the simple idea, these protocols combine two black-box consensus protocols thus are more complex.
Meanwhile, \framework combines the voting process in BFT consensus to Nakamoto-style consensus in a non-black-box way.
This greatly simplifies the protocol design: Nakamoto-style consensus is inherently liveness-favoring thus does not need sophisticated designs for ensuring liveness in BFT consensus (e.g., view change).

\section{Conclusion} \label{sec:conclusion}
We described \framework, a simple and performant framework for implementing BFT consensus for decentralized systems like blockchains.
\framework contains three protocols:
\sysnameSyn for synchronous networks,
and \sysnamePSyn and \sysname for partially synchronous networks.
Unlike existing BFT protocols, \framework adopts an egalitarian block production strategy, in which nodes randomly and noninteractively propose blocks with client transactions rather than relying on a leader to do so. \framework provides three features: no complicated fail-over protocols, better resilience to attacks on the leader, and comparable performance with state-of-the-art leader-based BFT protocols.

Through this work, we hope to raise awareness the egalitarian approach for simplifying BFT protocol design.
As well, our work reveals an intriguing connection between BFT protocols and Nakamoto-style consensus, which are usually regarded as two different families of Byzantine fault tolerant solutions.
We hope that our work sheds new light on this relationship and help researchers and developers to better understand these protocols and further formalize their inter-connections.

\noindent \textbf{Broadening view in BFT consensus design and implementation.}
Unlike existing leader-based BFT protocols~\cite{castro1999practical, kotla2007zyzzyva, HQReplica, 700BFT, HotStuffYin2019, shi2019streamlined, buchman2016tendermint}, \framework takes an egalitarian approach to simplifying the BFT consensus design.
This may seem counter-intuitive at first glance: it is a common belief that for Crash Fault Tolerant (CFT) consensus, leader-based protocols like Raft~\cite{raft2014} and Multi-Paxos~\cite{multipaxos} are simpler and more understandable than their egalitarian counterparts like Paxos~\cite{lamport2001paxos} and EPaxos~\cite{Epaxos}. So, why are leader-based BFT protocols much more complicated than their egalitarian version?
This is because leader-based BFT protocols have to introduce fail-over subprotocols to handle Byzantine behaviors that significantly increase the overall complexity.
Therefore, we advocate for an egalitarian approach to BFT protocol design.

More importantly, we illustrate the possibility of implementing BFT protocols on hundreds of time-tested blockchain platforms.
The lack of production-level systems hinders BFT consensus deployments, which has been pointed out by Bessani \etal~\cite{Bessani2014}: ``\textit{there are no robust-enough implementations of BFT SMR available---only prototypes used to validate novel ideas from papers---which makes it difficult to deploy this kind of technique in practice.''}
Our work indicates that we can use existing blockchains to implement BFT protocols with an egalitarian design.

\noindent \textbf{Making apple-to-apple comparison between BFT and NC possible.}
BFT protocols are believed to be quite different from Nakamoto-style consensus.
Due to these differences (stability favoring vs. egalitarianism favoring), it is often difficult to provide an apple-to-apple comparison between classic BFT consensus and Nakamoto-style consensus~\cite{BFTlens, lewis2020resource, lewis2021CCS, lewis2021byzantine}.
Our work sheds new light on what such a comparison could look like.
In particular, \framework adopts a  similar design to Nakamoto-style consensus, making a direct comparison not only possible, but also meaningful.
In this paper, we took the first step by implementing \framework on existing Nakamoto-style blockchains. In our future work we hope to perform a similar BFT augmentation with other blockchains.



\bibliographystyle{plain}
\bibliography{short}

\appendix

\section{Concentration Bounds}
In this section, we provide the concentration bounds that we use in the analysis. We denote the probability of an event $E$ by $\Pr[E]$ and the expected value of a random variable $X$ by $\e{X}$.

\begin{lemma}[Chernoff bounds for Poisson random variables]\label{lem:Poisson}
    Let $X$ be a Poisson random variable with mean $\mu$.  Then, for $0 < \delta < 1$, $\Pr\left( X \ge (1 + \delta) \mu \right) \le e^{-\delta^2 \mu / 3}$ and $\Pr\left( X \ge (1 - \delta) \mu \right) \le e^{-\delta^2 \mu / 2}$.
\end{lemma}

\begin{lemma} [Chernoff bound for dependent random variables \cite{niu2019analysis}] \label{lem:key_step}
    Let $T$ be a positive integer. Let $X^{(j)} = \sum_{i = 0}^{n-1} X_{j + iT}$ be the sum of $n$ independent indicator random variables and $\mu_j = \e{ X^{(j)} }$ for $j \in \{1, \ldots, T\}$. Let $X = X^{(1)} + \cdots + X^{(T)}$. Let $\mu = \min_j \{ \mu_j \}$. Then, for $0 < \delta < 1$, $\Pr\left[ X \le (1 - \delta) \mu T \right] \le e^{-\delta^2 \mu/2}$.
\end{lemma}

\section{Nakamoto Consensus} \label{sec:Nakamoto Code}
Algorithm~\ref{alg:Nakamoto} provides the pseudocode of Nakamoto consensus, which comprises three functions running in parallel. In particular, nodes use PoW to mine blocks after the longest chain that they have seen via $\textsf{MineBlock}$() function.
nodes use \textsf{SendMsg}() to continuously synchronize receiving blocks.
In the implementation, the block synchronization process can be realized by gossip protocols and the underlying P2P network~\cite{nakamoto2012bitcoin}.
When receiving new blocks, nodes use \textsf{ProcessBlock}() to verify blocks and update their blockchains.

\begin{algorithm}[t]
    \small
    \caption{The pseudocode of Nakamoto Consensus} \label{alg:Nakamoto}
    \begin{algorithmic}[1]
        \Statex \textbf{Local State}:
        \State $M \leftarrow \{\mathcal{G}_0 \}$ \textcolor{blue}{\Comment{the set of blocks}}
        \State $B^{\prime} \leftarrow \mathcal{G}_0$ \textcolor{blue}{\Comment{the highest block}}
        \Statex
        \State \textbf{procedure} $\textsf{MineBlock}()$
        \State\hspace{\algorithmicindent}
        $\textit{B.Txs} \leftarrow$ \textsf{getTransactions}()
        \State\hspace{\algorithmicindent}
        $\textit{B.hash} \leftarrow$ \textsf{H}($B^{\prime}$)
        \State\hspace{\algorithmicindent}
        $\textit{B.nonce} \leftarrow$ \textsf{getNewNonce}()
        \State\hspace{\algorithmicindent}
        \textbf{While} $\textsf{H}(B) > D$ \textbf{do} \textcolor{blue}{\Comment{the mining difficulty}}
        \State \hspace{\algorithmicindent} \hspace{\algorithmicindent}
        $\textit{B.nonce} \leftarrow$ \textsf{getNewNonce}()
        \State \hspace{\algorithmicindent}
        \textsf{ProcessBlock}($B$)
        \Statex
        \State \textbf{procedure} \textsf{SendMsg}()
        \State\hspace{\algorithmicindent}
        \textsf{Broadcast} $M$ \textcolor{blue}{\Comment{Using P2P network to synchronize missing blocks from neighbors}}
        \Statex
        \State \textbf{procedure} \textsf{RecvMsg}()
        \State \hspace{\algorithmicindent}
        \textbf{foreach} receiving block $B$ \textbf{do}
        \State \hspace{\algorithmicindent} \hspace{\algorithmicindent}
        \textsf{ProcessBlock}($B$)
        \Statex
        \State \textbf{procedure} \textsf{ProcessBlock}($B$)
        \State\hspace{\algorithmicindent}
        verify that \textsf{H}($B$) $<$ T
        \State\hspace{\algorithmicindent}
        verify that $\textit{B.hash} = \textsf{H}(A)$ for block $A \in M$
        \State\hspace{\algorithmicindent}
        verify that $\textit{B.Txs}$
        \State\hspace{\algorithmicindent}
        if (any of the above 3 verifications fails) \textbf{then return}
        \State\hspace{\algorithmicindent}
        $M \leftarrow M \cup \{B \}$
        \State \hspace{\algorithmicindent}
        $B^{\prime} \leftarrow \textsf{getHighestBlock}(M)$
    \end{algorithmic}
\end{algorithm}

\section{\sysnameSyn security analysis} \label{EBFT-Sync:security}

\subsection{Safety Analysis} \label{subsec:EBFT-Sync-safety}
The safety property guarantees that once a block is committed, there are no other committed blocks at the same height. In particular, a block is directly committed if a node commits it triggered by its timer, whereas a block is committed indirectly if a node commits it by its directly committed descendant block. To prove the safety property, we first present the following lemma.



\begin{lemma}\label{lem:synfinality}
    If an honest node directly commits $B_{k}$, then no conflicting block is certified at height $k$.
\end{lemma}

\begin{proof}
    Suppose an honest node directly commits $B_{k}$ at time $t$. Then, at time $t - 3\Delta$, the node has seen the longest certified chain extended by block $B_{k}$ (which is not certified yet).
    By the strong $\Delta$-bounded assumption of the synchronous network, this block $B_{k}$ together with its certified ancestor blocks will reach all honest nodes by time $t - 2\Delta$.
    Then, all honest nodes will send votes for this block unless some honest nodes have observed a certified chain with no less than $k$ length at time $t - 2\Delta$.
    If all honest nodes vote for this block at time $t - 2\Delta$, all nodes will receive at least $f+1$ votes by time $t - \Delta$ and then observe a certified chain ended with this certified block.
    By then, honest nodes will only vote for blocks with heights larger than $k$, and so $B_{k}$ is the only certified block at height $k$.
    Otherwise, if any node has voted for a conflicting block at height $k$ before $t - \Delta$, then any node will observe such a conflicting block with $B_{k}$ and does not commit $B_{k}$ at time $t$.
\end{proof}

This lemma says that a directly committed block is unique at its height.

\begin{theorem} [Safety]\label{synSafety}
    If blocks $B_{k}$ and $B^{\prime}_{k}$ at height $k$ are committed by some honest nodes, then $B_{k} = B^{\prime}_{k}$.
\end{theorem}

\begin{proof}
    Assume for contradiction that $B^{\prime}_{k} \neq B_{k}$ is committed by some honest nodes.
    (Note that the two blocks could be committed by a single node.)
    Suppose $B_k$ is committed as a result of directly committed $B_{\ell}$, and $B^{\prime}_{k}$ is committed as a result of directly committed $B_{h}$. This implies  $B_{\ell}$ extends $B_{k}$, and $B_{h}$ extends $B^{\prime}_{k}$. Without loss of generality, we assume $\ell \leq h$. By Lemma~\ref{lem:synfinality}, there is no certified block $B^{\prime}_{\ell}$ ($B^{\prime}_{k} \neq B_{\ell}$). Therefore, $B_{h}$ extends $B_{\ell}$, and $B^{\prime}_{k} = B_{k}$.
\end{proof}

\subsection{Liveness Analysis}\label{subsec:liveness}
The liveness property guarantees that client transactions will be eventually included in committed
blocks no matter what the adversary does.

\vspace{1mm} \noindent \textbf{Block producing model.} The block production of using cryptographic lottery can be modeled as a Poison process with the rate $\lambda$~\cite{niu2019analysis,UIUC,bagaria2019deconstructing,nakamotoRace}.
Let $\beta$ ($\beta < 1/2$) be the fraction of nodes controlled by the adversary.
Let $\lambda_a$ and $\lambda_h$ denote the block producing rate of the adversary and honest nodes, respectively.
In PoW, the probability that a block is produced by a node is proportional to its fraction of computation power.
Therefore, we have $\lambda_a = \beta \lambda$ and $\lambda_h = (1-\beta) \lambda$.
Besides, as $\beta < 1/2$, we have $\lambda_a < \lambda_h$.

\begin{figure}[t]
    \centering
    \includegraphics[width=2.6in]{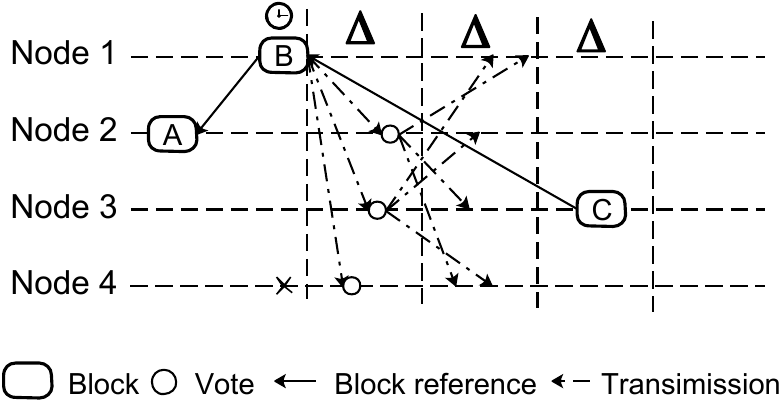}
    \setlength{\belowcaptionskip}{-0.5cm}
    \caption{\textbf{A simple case of uniquely certified blocks.} Block $B$ is a uniquely certified block since it is produced after the longest chains, and during its propagation and associated voting process, there are no other blocks.
    }
    \label{fig:certifiedBlock}
\end{figure}

\vspace{1mm} \noindent \textbf{Proof sketch.} Let us first revisit the block committing case in Figure~\ref{fig:certifiedBlock}, in which block $D$ is produced after (one of) the longest certified chain at time $t$, and in the next $2\Delta$ period there is no other blocks are produced.
Obviously, the block will become a uniquely certified block at its height.
Besides, if there is no adversarial block to match the block before $t + 3\Delta$, the block $D$ together with its non-committed ancestor blocks will be committed.
Hence, \sysnameSyn relies on unique certified blocks to reach finality.

In the following analysis, we first formalize a useful concept of \textit{uni-block}, i.e., an honest and uniquely certified block within a certain time period.
Then, we show that with high probability, there always exist such \textit{uniquely} certified blocks in a period $T$ no matter what the adversary does. The definition of uni-blocks is given below.

\begin{definition}[Uni-blocks]
    A block produced by honest nodes at time $t$ is called a uni-block if there is no honest block produced in the previous and next $2 \Delta$ time.
\end{definition}

Next, we prove that in a given interval $T$, there exist multiple uni-blocks (see Lemma~\ref{lemma:synnumber}). Before that, we first have the following two lemmas.

\begin{lemma}\label{lem:syn2delta}
    If an honest party has observed a block $B_{k}$ (not certified yet) that extends a certified blockchain at the time $t$, then every honest party can start to produce blocks of height at least $k+1$ by the time $t + 2\Delta$.
\end{lemma}

\begin{proof}
    First, this block $B_{k}$ and its certified ancestor blocks will reach all the honest nodes by time $t + \Delta$ by the $\Delta$-bounded assumption during periods of synchrony.
    If some honest nodes have observed other certified chains with no less than $k$ length at time $t + \Delta$, and do not vote for this block, then all nodes will observe this chain by time $t + 2\Delta$ (because of the $\Delta$-bounded assumption).
    Otherwise, all honest nodes will send votes for this block by LCCR, and so at least $2f+1$ votes will arrive at all honest nodes by time $t + 2 \Delta$. Then, all nodes will observe a certified chain ended with this certified block.
    Therefore, no matter in which cases, honest nodes will observe a certified chain with no less than $k$ length at time $t + 2\Delta$, and honest nodes start to produce blocks with a height of at least $k+1$.
\end{proof}

\begin{lemma} \label{lem:synUBlock}
    Suppose a block $B$ is a uni-block of height $k$, then no other honest block can be of height $k$.
\end{lemma}

\begin{proof}
    Suppose for contradiction that two honest blocks $B$ and $B'$ of height $k$ are produced at time $t$ and $t'$ respectively.
    Since no other honest block is produced between time $t-2\Delta$ and $t+2\Delta$, we have $t' \ge t+2\Delta$ or $t' \le t-2\Delta$. If $t' \ge t + 2\Delta$, by Lemma~\ref{lem:syn2delta}, every honest node observes a certified chain of length at least $k$ by time $t'$, and meanwhile, nodes produce  blocks on top of it.
    Therefore, no honest node will produce a new block of height $k$ after time $t'$, leading to a contradiction.
    Similarly, if $t' \le t-2\Delta$, every honest node observes a certified chain of length at least $k$ before the time $t$ (or even earlier), leading to a contradiction.
\end{proof}

With the above two lemmas, we can compute the bounded number of uni-blocks in an interval $T$.
\begin{lemma} \label{lemma:synnumber}
    Let $\eta = e^{-2 (1-\beta) \lambda \Delta}$. For any $0 < \delta <1$, the number of uni-blocks produced in a time interval $T$ is at least $(1+\delta) \eta^2  (1-\beta) \lambda T $, except for probability $e^{-\Omega\left(T\right)}$.
\end{lemma}

\begin{proof}
    Let $N_H(T)$ denote the number of honest blocks produced in a time interval $T$, and note that $\e{N_H(T)} = \lambda_h T = (1-\beta) \lambda T$.
    Then, for any $\delta_1 \in (0, 1)$, we have $\Pr[N_H(T) \leq (1 - \delta_1)(1-\beta) \lambda T] \leq e^{-\delta_1^2 (1-\beta) \lambda T / 2} = e^{-\Omega\left(T\right)}$ by Lemma~\ref{lem:Poisson}.
    In particular, let $k = (1 - \delta_1)(1-\beta) \lambda T$ be an integer by choosing a suitable $T$.
    We enumerate the first $k$ honest blocks produced since the start of the time interval as blocks $1, 2, . . . , k$.
    Without loss of generality, we assume there is a block $0$ (\resp block $k+1$) that is the last honest block produced before (\resp after) the interval.

    Let $X_i$ denote the block interval between $(i-1)$-th and $i$-th block. Recall that the block production process of honest nodes is the Poisson process with rate $(1-\beta) \lambda$.
    Hence, $X_i$ follows i.i.d. exponential distribution with the same rate.
    Let $Y_i$ denote an indicator random variable which equals one if the $i$-th block is uni-block and zero otherwise. Define $Y = \sum_{i=1}^{n}Y_i$.
    It is easy to see that the $i$-th block is a uni-block if $X_i \geq 2\Delta$ and $X_{i+1} \geq 2\Delta$. Since $X_i$ and $X_{i+1}$
    are independent, we have $\Pr[Y_i = 1] = \Pr[X_i \geq 2\Delta]\Pr[X_{i+1} \geq 2\Delta] = e^{-4 (1-\beta) \lambda \Delta} = \eta^2$.
    Note that $Y_i$ and $Y_{i+1}$ are not independent since they both depend on the event that $X_{i+1} \geq 2\Delta$, but $Y_i$ and $Y_{i+2}$ are independent.
    Thus, $Y$ can be broken up into two summations of independent Boolean random variables $Y = \sum_{odd}{Y_i} + \sum_{even}{Y_i}$.
    By Lemma~\ref{lem:key_step}, we have $\Pr[Y \leq (1-\delta) \eta^2 (1-\beta) \lambda T] \leq e^{-\delta^2 \eta^2 (1-\beta) \lambda T/2} = e^{-\Omega\left(T \right)} $.
\end{proof}

By Lemma~\ref{lemma:synnumber}, we can prove that under some conditions, the adversary cannot produce conflicting blocks to match each uni-blocks in the following lemma.

\begin{lemma} \label{lemma:synlivess1}
    Suppose $\eta^2 (1-\beta) > (1+\delta) \beta$. In a time interval $T$, there exist uni-blocks in the longest chain, except for $e^{-\Omega\left(T\right)}$ probability.
\end{lemma}

\begin{proof}
    Let $N_H(T)$ (\resp $N_A(T)$) denote the number of uni-blocks (\resp adversarial block) produced in the interval $T$.
    By Lemma~\ref{lemma:synnumber}, the number of uni-blocks is $N_H(T) > (1-\delta_1) \eta^2 (1-\beta) f T$ except for $e^{-\Omega\left(T\right)}$ probability.
    Similarly, the expected time for the adversary to produce a block is $\frac{1}{\lambda_a}$. In the best case, the adversary can immediately transmit a block to honest nodes and then obtain their votes without delay.
    This means that the adversary can immediately produce the next block on its new block.
    Thus, during a time interval $T$, the adversary can produce blocks at most $(1+\delta_2)\lambda_a T$ except for probability $e^{-\delta_2^2 \lambda_a T/3} = e^{-\Omega(T)}$ by Lemma~\ref{lem:Poisson}.

    By setting $\delta_1 = \delta_2 = \delta/4$ and noticing $\frac{1 + \delta/4}{1 - \delta/4} < 1 + \delta$, we have
    $(1 - \delta_1) \eta^2 (1-\beta) > (1 + \delta_2) \beta$.
    Therefore, $N_H(T) > N_A(T)$ except for $e^{-\Omega\left(T\right)}$ probability.
    This means that the adversary cannot create conflicting blocks to match each uni-blocks.
\end{proof}

\begin{lemma} \label{lemma:step}
    If two certified blocks $B$ and $B^{\prime}$ are observed by some honest nodes, there must exist an honest node that has voted for both of them.
\end{lemma}

We next prove the liveness property of \sysnameSyn.
\begin{theorem}[Liveness]\label{theo:synLiveness}
    Suppose $\eta^2 (1-\beta) > (1+\delta) \beta$. In a time interval $t$, there exist committed honest blocks in the main chain except for $e^{-\Omega\left(T \right)}$ probability.
\end{theorem}

\begin{proof}
    By Lemma~\ref{lemma:synlivess1}, there exists at least one uni-blocks in the longest chain except for $e^{-\Omega\left(T \right)}$ probability.
    Without loss of generality, we assume that such a block $B$ is produced at time $t$.
    By time $t + 3 \Delta$, all nodes will commit this block together with its non-committed ancestor blocks.
\end{proof}

\begin{remark}
    The above liveness analysis is loose because of the assumption of a very powerful adversary. The adversary can control the lottery winning timing, and use each winning chance to produce blocks that conflict with uniquely certified blocks by honest nodes. However, the adversary cannot do this through a cryptographic lottery. We leave more tight analysis as one of our future work.
\end{remark}

\subsection{Adaptive security}\label{subsubsec:ebft-syn-adaptive-sec}
The adaptive corruption does not affect safety or liveness of \sysnameSyn.
For safety, the adaptive corruption does not affect quorum intersection.
For liveness, the adaptive corruption does not affect the uni-blocks' distribution which is secured by the lottery, or certifying blocks which ensures a quorum number of votes over uni-blocks.

\section{\sysnamePSyn Security Analysis} \label{EBFT-PSync:security}

\subsection{Safety Analysis}\label{subsec:safety}
The safety property guarantees that there are no conflicting committed blocks at the same height under any network conditions. Similarly, a block $B_k$ is directly committed if its direct child block $B_{k+1}$ contains at least $2f+1$ \comVotes. Otherwise, a block is indirectly committed.
We first present one useful lemma of directly committed blocks.

\textit{Lemma~\ref{lemma:Psystep1}:} If an honest node directly commits $B_{k}$, then no conflicting block is certified at height $k$.

\begin{proof}
    Let $\mathcal{S}$ (resp, $\mathcal{S}^{\prime}$) denote the set of honest nodes that have voted for block $B$ (\resp $B^{\prime}$). As there are at most $f$ Byzantine nodes (who can vote twice for both block $B$ and $B^{\prime}$), set $\mathcal{S}$ and $\mathcal{S}^{\prime}$ at least contain $f+1$ honest nodes. The intersection of these two sets (of honest nodes) is $|\mathcal{S} \cap  \mathcal{S}^{\prime}| \geq (f+1) + (f+1) - (2f+1) = 1 > 0$. This implies that at least one honest node must have voted for both block $B$ and $B^{\prime}$.
\end{proof}

\begin{lemma} \label{lemma:Psystep1}
    If an honest node directly commits $B_{k}$, then no conflicting block is certified at height $k$.
\end{lemma}

\begin{proof}
    First, these must exist a certified chain (seen by the honest node) that ended with two blocks $B_{k}$, $B_{k+1}$ at height $k$ and $k+1$, and block $B_{k+1}$'s QC contains no less than $2f+1$ \comVotes. Assume for contradiction that $B^{\prime}_{k} \neq B_{k}$ is certified in at least one honest node, as shown in Figure~\ref{fig:safety}.
    By Lemma~\ref{lemma:step}, there must exist one honest node that has voted for these two blocks.
    In particular, this node has sent a \comVote for block $B_{k+1}$.
    Let $t$ (\resp $t_1$) denote the time when this node voted for block $B_{k+1}$ (resp, $B^{\prime}_{k}$).
    No matter $t < t_1$ or $t > t_1$, we will drive a contradiction. (Here, a node is assumed to sequentially vote for blocks, so we do not consider the case $t = t_1$.)

    \begin{packeditemize}
        \item $t < t_1$. The honest node first voted for block $B_{k+1}$. According to the voting rule, when the node receives block $B^{\prime}_{k}$, it would not vote for it. This is because the node has seen a certified chain of length $k$, and obviously, $B^{\prime}_{k}$ does not satisfy LCCR. This leads to a contradiction.

        \item $t > t_1$. The honest node first voted for block $B^{\prime}_{k}$. When the node later received $B_{k+1}$, it would send a \witVote for it because the block $B^{\prime}_{k}$ has the same height as block $B_{k+1}$'s parent block $B_{k}$, and this node has already voted it. This leads to a contradiction that this node has sent a \comVote for block $B_{k+1}$.
    \end{packeditemize}
\end{proof}

\textbf{\begin{figure}[t]
        \centering
        \includegraphics[width=2.3in]{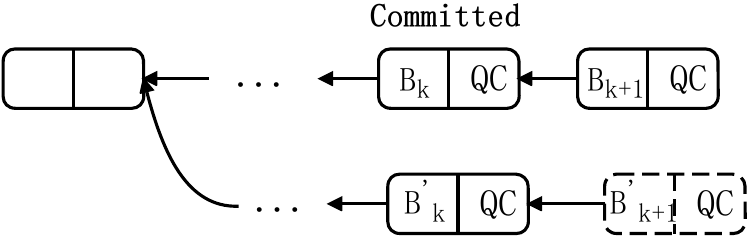}
        \caption{\textbf{The safety violation cases (impossible).} It is impossible to have a certified block $B_{k}^{\prime}$ and a committed block $B_{k}$ at height $k$. Since block $B_{k}$ is directly committed by its child block $B_{k+1}$ who has at least $2f+1$ \comVotes.  }
        \label{fig:safety}
    \end{figure}}

With this lemma, the following theorem will show that the safety property holds for all committed blocks.

\begin{theorem} [Safety]
    If blocks $B_{k}$ and $B^{\prime}_{k}$ at height $k$ are committed by some honest nodes, then $B_{k} = B^{\prime}_{k}$.
\end{theorem}

\begin{proof}
    Assume for contradiction that $B^{\prime}_{k} \neq B_{k}$ is committed by some honest nodes.
    (Note that the two blocks could also be committed by a single node.)
    Suppose the block $B_{k}$ is committed as a result of certified blocks $B_{v}$ and $B_{v+1}$, and $B^{\prime}_{k}$ is committed as a result of  certified blocks $B_{\ell}$ and $B_{\ell+1}$. Both blocks $B_{v+1}$ and $B_{\ell+1}$ contain no less $2f+1$ \comVotes.
    Clearly, we have $v, m \geq k$.
    Without loss of generality, we assume that $v \leq \ell$.
    By Lemma~\ref{lemma:Psystep1}, there is no certified block $B^{\prime}_{v}$ ($B^{\prime}_{v} \neq B_{v}$). Therefore, $B_{\ell}$ extends $B_v$ and $B_k = B^{\prime}_k$.
\end{proof}

\subsection{Liveness Proof}
The liveness property guarantees that honest blocks (including client transactions) will be eventually committed, no matter what the adversary does. Specifically, the liveness property holds only when the network is synchronous, \ie, after the \textsf{GST}.
This is because when the network is partitioned and delays can be arbitrarily long, no certified blocks can be produced (without enough votes), and consequently, no blocks can be committed.

\vspace{1mm} \noindent \textbf{Proof sketch.}
\sysnamePSyn also relies on uni-blocks to realize finality.
Informally speaking, if there exists a uni-block, and the block remains unique at its height until it is later extended by a certified descendant block (\ie, with at least $2f+1$ \comVotes), the block together with its ancestor blocks will be committed.
To this end, we first prove that the liveness property holds when $\textsf{GST} = 0$.
The proof of this case is the same as that in the synchronous network, and has already been proved in Theorem~\ref{theo:synLiveness}.
Next, we extend the above proof by making $\textsf{GST} > 0$.
The difference in the liveness proof is that the adversary can withhold some blocks before \textsf{GST} due to the asynchronous network. However, the next analysis shows that the difference does not affect the existence of uni-blocks in a time interval of $T$.

\vspace{1mm} \noindent \textbf{Detailed analysis.}
We first have the following lemmas, proving that the adversary can only hide a finite number of blocks that are higher than the highest block that any honest nodes know after $\textsf{GST} + 2\Delta$.

\begin{lemma}[Bounded number of hidden blocks]\label{lem:withholding}
    At any time $t \geq \textsf{GST} + 2\Delta$, the number of unknown blocks to any honest nodes is bounded with high probability.
\end{lemma}

\begin{proof}
    Without loss of generality, we assume that the block $B_{k}$ is the highest block published by the adversary before the time $t - 2\Delta$.
    Specifically, to produce the block $B_{k}$, all ancestor blocks of the block $B_{k}$ have been certified.
    This implies, at least $f+1$ honest nodes have seen and voted for block $B_{k}$'s parent block before the time $t - 2\Delta$.
    (In the ideal case, we assume that the adversary immediately collects all votes for block $B$'s parent block, and then generates the block $B$ without delay.)
    By Lemma~\ref{lem:syn2delta}, all honest nodes would observe a certified chain with no less than $k$ length by time $t$.
    Let $N_A(2\Delta)$ denote the number of produced adversarial blocks between the time $t - 2\Delta$ and time $t$.
    As the block producing process of the adversary is a Poisson process with the rate $\lambda_h$, the probability of generating $k$ new blocks that extend the block $B_{\ell}$ is $\frac{e^{-2\lambda_a\Delta} (2\lambda_a\Delta)^K}{k!}$, which drops exponentially with the increase of $k$.
\end{proof}

This lemma implies that any honest nodes do not know a finite number of blocks that are higher than the highest block that they have known. Therefore, by increasing the interval $T$, \sysnamePSyn can guarantee that these exist certified unique blocks, which will be committed with high probability. This establishes the liveness of \sysnamePSyn.

\begin{theorem}[Liveness]
    Suppose $\eta^2 (1-\beta) > (1+\delta) \beta$. In a time interval $T$, there exist committed honest blocks in the main chain except for $e^{-\Omega\left(T \right)}$ probability.
\end{theorem}

\begin{proof}
    First, by Lemma~\ref{lem:withholding}, the adversary can only withhold a finite number of blocks.
    Once the network is synchronous (\ie, $t > \textsf{GST}$), \sysnamePSyn can guarantee liveness by Theorem~\ref{theo:synLiveness}.
\end{proof}

\subsection{Adaptive security}\label{subsubsec:ebft-psyn-adaptive-sec}
Similar to \sysnameSyn, adaptive corruption does not affect safety or liveness of \sysnamePSyn.
For safety, adaptive corruption does not affect quorum intersection.
For liveness, adaptive corruption does not affect the distribution of uni-blocks or the voting process.

\end{document}